\documentclass[a4paper,11pt]{article}
\usepackage{jheppub}	
\usepackage[utf8]{inputenc}
\usepackage[english]{babel}
\usepackage{makeidx}
\usepackage{amsfonts}
\usepackage{enumerate}
\usepackage{mathrsfs}
\usepackage{tensor}
\usepackage[autostyle]{csquotes}
\usepackage{subfig}
\usepackage{simpler-wick}

\newdimen\tableauside\tableauside=1.0ex
\newdimen\tableaurule\tableaurule=0.4pt
\newdimen\tableaustep
\def\phantomhrule#1{\hbox{\vbox to0pt{\hrule height\tableaurule
width#1\vss}}}
\def\phantomvrule#1{\vbox{\hbox to0pt{\vrule width\tableaurule
height#1\hss}}}
\def\sqr{\vbox{%
  \phantomhrule\tableaustep
\hbox{\phantomvrule\tableaustep\kern\tableaustep\phantomvrule\tableaustep}%
  \hbox{\vbox{\phantomhrule\tableauside}\kern-\tableaurule}}}
\def\squares#1{\hbox{\count0=#1\noindent\loop\sqr
  \advance\count0 by-1 \ifnum\count0>0\repeat}}
\def\tableau#1{\vcenter{\offinterlineskip
  \tableaustep=\tableauside\advance\tableaustep by-\tableaurule
  \kern\normallineskip\hbox
    {\kern\normallineskip\vbox
      {\gettableau#1 0 }%
     \kern\normallineskip\kern\tableaurule}%
  \kern\normallineskip\kern\tableaurule}}
\def\gettableau#1 {\ifnum#1=0\let\next=\null\else
  \squares{#1}\let\next=\gettableau\fi\next}

\def\a{\alpha}		\def\b{\beta}				\def\d{\delta}
						
				\def\l{\lambda}		
									\def\r{\rho}	
\def\s{\sigma}		\def\t{\tau}						
\def\c{\chi}

\def\be{\begin{equation}}
\def\ee{\end{equation}}
\def\bea{\begin{eqnarray}}
\def\eea{\end{eqnarray}}

\title{ Euclidean Time Approach to Entanglement Entropy on Lattices and Fuzzy Spaces}

\author{A. Allouche,}
\author{D. Dou}

\affiliation{Dept of Physics, College of Exact Sciences, Hamma Lakhdar University, El Oued, Algeria.}
\affiliation{Lab of Linear Operators and Partial Diff Equations, Theory and Application, Hamma Lakhdar Univeristy.}

\emailAdd{dou-djamel@univ-eloued.dz}

\abstract{In a recent letter, \emph{Physics Letters B 792 (2019) 60–63}, we developed a novel  Euclidean time approach to compute R\'{e}nyi entanglement entropy on lattices and fuzzy spaces based on Green's function.  The present work is devoted in part to the  explicit proof of the Green's matrix function formula  which was  quoted and used in the previous letter, and on the other part to some applications of this formalism. We focus on  scalar theory on 1+1 lattice. We also use the developed approach to systematically go   beyond the Gaussian case by considering interacting models,  in particular our results confirm  earlier expectations  concerning the correction to the entanglement at first order. We finally outline how this approach can be used to compute the entanglement entropy on fuzzy spaces for free and interacting scalar theories}

\keywords{Entanglement Entropy, Green's Function, Euclidean Time, Black holes, Lattices, Fussy Spaces}

\preprint{PREPRINT}
 \clearpage

\begin{document}
\maketitle


\section{Introduction}

Quantum entanglement  plays  fundamental roles in different areas of physics and future technology, it is generic and ubiquitous, starting from the role it plays in quantum information, being  primary resource for quantum computation and information
 processing,   to  condensed matter, black holes  and  high energy physics, see  \cite{Nishioka:2018khk, Plenio:2007zz, Headrick:2019eth, Calabrese:2004eu},  .

 Moreover, in view of several recent development and progress made in connection with holography and strongly-couple quantum field theories that have led to some inspirational ideas in defining quantum gravity from quantum many-body entangled states, it seems that quantum entanglement and holography will be two major pieces of our understanding of the fundamental nature of space-time.

 Many measures have been developed to quantify   the degree of entanglement between different parts of a quantum system \cite{Nishioka:2018khk, Casini:2009sr, Calabrese:2004eu}.  One of the most important and fruitful measures is
 of course the entanglement entropy (EE).

 In connection with black hole entropy and
 quantum field theory the finding and observations  made in \cite{Sorkin:1984kjy, Bombelli:1986rw}   led subsequently to a wide interest  in  EE   in the presence of black hole, as one would have to expect some amount of entropy to accompany an event horizon, since it is by definition an information hider.

 Several techniques have been devised to compute this entropy  depending on the area
 of research field. Generally those techniques fall into two categories. Real time and Euclidean time formalisms \cite{Nishioka:2018khk}.

 The real time approach  was initially developed in \cite{Bombelli:1986rw, Srednicki:1993im}, during the quiet period of the subject, and later reformulated with deeper insights using the two point correlators of the field variables \cite{Peschel2003}. This formalism is very suitable for numerical computation once the underlying space is discretized to lattice or for a bosonic or fermionic chain, however this approach seems to be  limited to gaussian states or free field theories.

 The results  of \cite{Bombelli:1986rw, Srednicki:1993im} trigged  attention to the EE in the black hole context and soon an Euclidean time   approach was   developed in \cite{Callan:1994py, Kabat:1994vj}  based on   Euclidean heat kernel
 and Green's function. This formalism proved to be more powerful  in performing analytical computation in QFT and extracting the UV divergence. Moreover it allows one to make connection with string theory \cite{Susskind:1994sm},
 extract finite  mass dependent but cuttoff independent  contribution to the area law \cite{Hertzberg:2010uv}. Unlike the real time approach the Green's function approach can be used   to go
 beyond the free case and  investigate the  dependence of entanglement  entropy on the renormalized mass
 \cite{Hertzberg:2012mn}.

For  $1+1$  QFT,  the tools of CFT were  very early employed to investigate the EE \cite{Holzhey:1994we}, ever since  a lot of work and substential advances have been made in the understanding of  EE in 1+1, see \cite{Calabrese:2009qy, Headrick:2019eth} and references therein.

 Now, although the heat kernel and the Green's function approaches offer  a powerful tool to compute and
 investigate the entanglement entropy they seem to be limited to some special cases of continuum
 geometries.  For example when one traces out half of flat $d+1$ spacetime   R\'{e}nyi entropy turns out
 to be governed by  a simple geometry, namely a cone with deficit angle $2\pi (1-n)$, for which the
 Green's function is known or the heat kernel expansion can be made. However, if one traces out a finite
 region it becomes almost untractable problem to construct the corresponding Green's function or to
 recognize the resulting $n$-sheeted  geometry.

 Motivated by the absence of an Euclidean formalism for the EE on lattices and fuzzy spaces  and the limitations of the real time approach  to free and gaussian states   an  Euclidean time approach     based on Green's function  has recently been proposed in  \cite{Allouche:2018err}. An alternative formula to compute R\'{e}nyi EE resulting from tracing out an arbitrary set of intervals  was obtained for a one-dimensional chain of coupled H.O's.

 By its very formulation the proposed approach furnishes a setting on which a systematic  perturbation series for the EE in interacting theories can be obtained.

 The  fundamental piece upon which  this approach rests  is the Green's matrix function which was quoted  in \cite{Allouche:2018err} and the  proof of its construction was deferred to the present work. In this paper we  show how the Green's  function is constructed for an arbitrary set of intervals. We further   apply the formalism to a scalar field theory on 1+1 lattice by considering the EE in two important particular cases, one resulting from tracing out half of the space and a second resulting from tracing out an interval of infinitesimal size, in particular an interval made of a single  point. For both cases we perform analytical study.

 From the technical point of view  the case of one-point  turned out to be easy to handle analytically within this approach and we could recover explicitly the formalism developed in \cite{Peschel2003}. However, the case of half-space EE necessitated  the finding of an asymptotic inverse for a special  Toeplitz matrix with N-dependent symbol. This problem   turned out to be a strenuous one due, as far as we are aware, to the lack of any solution for this kind of problem in mathematical literature.  Therefore we hope that our  modest contribution will pave the way to a better  understanding of this class of Toeplitz matrices.

 As another important application of the present technique we consider the correction to the EE due to a quartic type interaction for  a system made of finite number of coupled oscillators. In particular we derive explicitly the first order correction to the  standard EE between two coupled harmonic oscillators due to quartic interaction, $\lambda\sum q_i^4$, and show that to the first order in perturbation theory the standard gaussian (free) EE formula stays  valid for interacting theories ( non-gaussian)   with a shifted parameter due to the perturbation of the ground state. Our result systematically  confirm earlier conjecture and expectations that the EE is generally a space-time   quantity and  captured by the perturbed two-points  correlators to the first order in perturbation theory \cite{Chen:2020ild} .

 In addition to the above quartic interaction, to which we shall refer as of local origin, we consider another natural quartic interaction which can not be seen as arising from a discretization of  a non-local term. In contrast to the local interaction this quartic interaction appears to have  an IR divergence that can not be normalized away and seems to limit the type of quartic interactions that the entanglement can tolerate .

 The present work is organized as follows. In the first section we begin by a subsection in which we review the formalism developed in \cite{Allouche:2018err}. The second subsection will be devoted to   the detail of the construction of Green's matrix function. In section II we consider a massive scalar theory in 1+1 lattice and consider the EE for two cases, half of the space and one-point,  and as  a by-product we obtain the standard formula of the EE for two coupled H.Os.  A subsection is  devoted to the mathematical problem of finding an asymptotic inverse of a particular class of Teoplitz matrices with N-dependent index.  In  the third section we  go beyond the gaussian states  by including  quartic interaction and derive a general formula for computing the first order correction in perturbation theory. We give explicit first order evaluation of the correction to the EE for 2-coupled H.O and show that for a  typical quartic interaction the non-gaussian formula for  EE keeps the same form as the standard gaussian one and is captured by the perturbed two-point correlators. We conclude our paper by an outlook on the applications of the developped formalism to free and interacting scalar field theories on fuzzy spaces.

 Most of the technical detail of the paper are moved to  the appendices section.

 \section{The Green's Function Matrix for Arbitrary Set}

 To make this paper as self-contained as possible   we start by reviewing the formalism developed in \cite{Allouche:2018err}.

 \subsection{The Formalism}
 Consider a system of $N$-coupled H.O's  with standard quadratic   Lagrangian

 \be\label{LagStand}
 L= \frac{1}{2}( \dot{Q}^T \dot{Q} -Q^T V Q),~~ ~~~ Q=\left(
 \begin{array}{c}
 	q_1 \\
 	q_2 \\
 	\vdots\\
 	
 	q_N \\
 	
 \end{array}
 \right)
 \ee

 $V$ is a symmetric positive definite $N \times N$ matrix.

 By considering $ N$ coupled H.O
 we cover a wide class of field theories on lattices and fuzzy spaces. This follows from the observation  that many
 physical situations  are modeled by a chain of coupled H.O's ( Fermionic and Bosonic),
 whereas in high energy physics it turns out that for most theories regularized on lattices and fuzzy spaces
 computing the entanglement entropy reduces to considering  independent sectors made  of coupled H.Os, and this is independently of the dimension of the spacetime considered and the geometry of the region traced out  \cite{Nishioka:2018khk,Srednicki:1993im, Dou:2006ni, Dou:2009cw}

 Let now  $\rho=|0><0|$  be the ground state of whole system and  consider the reduced  density operator
 $\rho_A= \mathrm{Tr}_A \rho$ resulting from tracing out a subset $A$ of the field variables $Q$. $A$  could be  a union of collection of disjoint subsets . Let $\bar{A}$ denote the complimentary subset.

 For pedagogical reasons we shall consider the important special case where  $A=\{q_{p+1}, q_{p+2},.....q_{N}\}$. It will become clear as we move on that the proof of the final general result  is not tied to this particular choice.

 Let  $Q_A\equiv(q_{p+1}, q_{p+2},.....q_{N})^T$  and
 $Q_{\bar{A}}\equiv(q_{1}, q_{1},.....q_{p})^T$.

 The density matrix elements of the ground state can be written  using Euclidean path integral representation as

 \be
 \r (Q'_{\bar{A}},Q'_{A} ;Q''_{\bar{A}},Q''_A) =\int \mathcal{D}Q \exp
 (-\int_{-\infty(Q=0)}^{0^-(Q=(Q'_{\bar{A}},Q_{A}'))}L_E
 d\t-\int_{0^+(Q=(Q''_{\bar{A}},Q''_{A}))}^{\infty(Q=0 )}L_Ed\t)
 \ee

 The reduced density matrix is then given by,
 \be
 \r_A (Q'_{\bar{A}} ;Q''_{\bar{A}})=\int  \r (Q'_{\bar{A}},Q_{A}
 ;Q''_{\bar{A}},Q_A)dQ_A,~~~~~dQ_A=\prod_{A}dq_i
 \ee

 from which it follows that

 \be
 \mathrm{Tr}_{\bar{A}}\r_A^n=\int \r_A (Q^1_{\bar{A}} ;Q^2_{\bar{A}})\r_A (Q^2_{\bar{A}}
 ;Q^3_{\bar{A}})\cdot \cdot \cdot \cdot\r_A (Q^{n}_{\bar{A}} ;Q^1_{\bar{A}}) dQ^1_{\bar{A}}
 dQ^2_{\bar{A}}\cdot\cdot\cdot dQ^{n}_{\bar{A}}
 \ee

 Now let us introduce an $n$-fold field variable $\mathbb{Q}$ with $Nn$ components and $n$-fold potential $\mathbb{V}$ as
 \be
 \mathbb{Q}= \left(
 \begin{array}{c}
 	Q^1_{\bar{A}} \\
 	Q^1_A \\
 	Q^2_{\bar{A}}  \\
 	Q^2_A \\
 	\cdot\\
 	\cdot\\
 	\cdot\\
 	
 	Q^n_{\bar{A}} \\
 	Q^n_{A}
 	
 \end{array}
 \right)
 ~~   \mathrm{and}~~~\mathbb{V}=
 \mathbb{I}_{n\times n}  \otimes  V
 \ee

 then it is straightforward to write
 \bea\label{pi1}
 \mathrm{Tr}_{\bar{A}}\r_A^n &=&\int \mathcal{D}\mathbb{Q} \exp
 (-\int_{-\infty,\mathbb{Q}=0}^{0^-,\mathbb{Q}=(Q^1_{\bar{A}},Q^1_{A};Q^2_{\bar{A}},Q^2_{A}\cdot\cdot\cdot
 	Q^n_{\bar{A}},Q^n_{A} )}\mathbb{L}_E
 d\t \nonumber\\
 &-&\int_{0^+,\mathbb{Q}=(Q^2_{\bar{A}},Q^1_{A};Q^3_{\bar{A}},Q^1_{A}\cdot\cdot\cdot
 	Q^n_{\bar{A}},Q^{n-1}_{A};Q^1_{\bar{A}},Q^{n}_{A})}^{\infty, \mathbb{Q}=0}\mathbb{L}_E d\t)
 \eea

 where $\mathbb{L}_E$ is given by
 \be
 \mathbb{L}_E=\frac{1}{2}( \dot{\mathbb{Q}}^T \dot{\mathbb{Q}} +\mathbb{Q}^T\mathbb{ V} \mathbb{Q})
 \ee

 Unlike the continuum flat case we can not interpret the above path integral as  the partition
 function  on the $n$-fold cover manifold (with a conical singularity)  of the original manifold. However, it turns out that it is still possible to write  $\mathrm{Tr}_{\bar{A}}\r_A^n$ as a zero temperature partition function with
 specified Lagrangian.

 Let us note that the main obstacle which is preventing us from interpreting (\ref{pi1}) as a zero temperature partition function is the boundary conditions on the  variables $\mathbb{Q}_{\bar{A}}$ at the cut.  This can be surmounted by moving the cut, or the discontinuity, from  the integration variables to the potential via a permutation matrix $ \mathbb{P}_\pi$ which maps
 $(Q^1_{\bar{A}},Q^1_{A};Q^2_{\bar{A}},Q^2_{A}\cdot\cdot\cdot Q^n_{\bar{A}},Q^n_{A} )$ into
 $(Q^2_{\bar{A}},Q^1_{A};Q^3_{\bar{A}},Q^2_{A};\cdot\cdot\cdot
 Q^n_{\bar{A}},Q^{n-1}_{A};Q^1_{\bar{A}},Q^{n}_{A})$.

 Then $ \mathrm{Tr}_{\bar{A}}\r_A^n$  with the correct normalization can be written as
 \be\label{partionf}
 \mathrm{Tr}_{\bar{A}}\r_A^n = \frac{\mathbb{Z}_n}{\mathbb{Z}_1^n}=\frac{\int \mathcal{D} \mathbb{Q}
 	e^{-\mathbb{S}_E}}{(\int \mathcal{D}Q e^{-S_E})^n}
 \ee
 where
 \be\label{action}
 \mathbb{S}_E=\int_{-\infty}^{\infty} ( \frac{1}{2}[ \dot{\mathbb{Q}}^T \dot{\mathbb{Q}}
 +\mathbb{Q}^T\mathbb{ V(\t)} \mathbb{Q}] d\t, ~~~\mathbb{V} (\t)= \theta(-\t) \mathbb{V}+ \theta(\t)
 \mathbb{V}_p
 \ee

 it  follows immediately that

 \be\label{id}
 \ln \mathrm{Tr} \rho_A^n= -\frac{1}{2} \ln \frac{\det (-\frac{d^2}{d\t^2}+\mathbb{V}(\t))}{\det
 	(-\frac{d^2}{d\t^2}+V)^n},~~~~-\infty <\t < \infty
 \ee

 where $ \mathbb{V}(\t)$ is a $n$-fold step-potential given by
 \be
 \mathbb{V} (\t)= \theta(-\t) \mathbb{V}+ \theta(\t) \mathbb{V}_p
 \ee

 $\mathbb{V}$ and its image $\mathbb{V}_p$ are $Nn\times Nn $ matrices defined as follows
 \be
 \mathbb{V}= \mathbb{I}_{n\times n}  \otimes  V ~~  \mathrm{and} ~~  \mathbb{V}_p= \mathbb{P}_\pi
 \mathbb{V}\mathbb{P}_\pi^T
 \ee

 $\mathbb{P}_\pi$ is  $Nn\times Nn $ permutation matrix, its explicit form is given for the above particular choice of $A$  by
 \be
 \mathbb{P}_{\pi}(A,\bar{A})=
 \left(
 \begin{array}{cccccc}
 	P_{A}& P_{\bar{A}} & 0_{2N}&  \dots & 0_{2N}&0_{2N}\\
 	0_{2N}& P_{A}&P_{\bar{A}}&  \dots  & 0_{2N}& 0_{2N}\\
 	0_{2N}& 0_{2N}& P_{A}& \dots & 0_{2N}& 0_{2N}\\
 	\vdots& \vdots& \vdots&\ddots & \vdots& \vdots\\
 	0_{2N}& 0_{2N}& 0_{2N}& \dots & P_{A}& P_{\bar{A}}\\
 	P_{\bar{A}}& 0_{2N}& 0_{2N}& \dots & 0_{2N}& P_{A}
 \end{array}
 \right)
 \ee

 $P_{A}$ and $P_{\bar{A}}$ are  the projectors on $A$ and $\bar{A}$ resp. More explicitly we have

 $$
 P_{A}= \left(
 \begin{array}{cc}
 0_p & 0\\
 0&I_{N-p}
 \end{array}
 \right)
 ,~~~~ P_{\bar{A}}= \left(
 \begin{array}{cc}
 I_p & 0\\
 0& 0_{N-p}
 \end{array}
 \right)
 $$

 Actually it is easy to see that all permutations matrices for this kind of problems are made of projectors on the available and the unavailable degrees of freedom and zeros.

 If we decompose  the potential accordingly  as

 \be
 V= \left(
 \begin{array}{cc}
 	A & B \\
 	B^T & C \\
 \end{array}
 \right)
 \ee
 $A$ and $C$ are $p\times p$ and $(N-p)\times(N-p)$ matrices resp.

 The $n$-fold potential is  given for arbitrary $n$ by

 \be\label{matrix}
 \mathbb{V}_p= \mathbb{P}_\pi
 \mathbb{V}\mathbb{P}_\pi^T=
 \left(
 \begin{array}{cccccccc}
 	A & 0 & 0 & 0& 0&  \dots  & 0& B \\
 	0 & C & B^T &0 & 0 & \dots  & 0& 0 \\
 	0 & B &A & 0 & 0 & \dots & 0 & 0\\
 	0 & 0 & 0 & C & B^T & \dots & 0& 0\\
 	0 & 0 & 0 & B & A & \dots & 0& 0\\
 	\vdots & \vdots & \vdots & \vdots & \ddots & \vdots& A& 0 \\
 	B^T& 0 & 0 & 0 & \dots & 0 & 0&  C
 \end{array}
 \right)
 \ee

 The form of this supermatrix is typical in the calculation of the entanglement entropy using Euclidean formalism \cite{Callan:1994py, Shiba:2014uia}, and here it is understood as arising from the action of a special permutation  matrix.

 Now, as mentioned earlier, although the derivation of (\ref{id}) was explicitly made for a particular choice of the subset $A$, it remains valid for any other choice, one has to only change the permutation matrix. Therefore each tracing operation is characterized by a its permutation matrix  $\mathbb{P}_\pi (A,\bar{A})$, and all we  need is its action on $\mathbb{V}$ or the image potential $\mathbb{V}_p$.

 \subsection{Constructing Green's Matrix Function }

 This section is devoted to the proof of the main formulas which appeared in \cite{Allouche:2018err}, namely the explicit construction of the Green's function associated with this type of problems.

 We start by expressing (\ref{id}) using the heat Kernel matrix associated with the operator  $\hat{\mathbb{ A}}_n=-\frac{d^2}{d\t^2}+\mathbb{V}(\t)$.

 Using the eigenvalues and eigenfunctions  of $\hat{\mathbb{ A}}_n$ we can write
 \be\label{heategf}
 \mathbb{ K}_n(\t,\t',s)= \sum_{l} e^{-\lambda_l s} \Phi_l(\t)\Phi_l^{\dagger}(\t'),~~~~~~ \hat{\mathbb{ A}}_n \Phi_l(\t)=\l_l \Phi_l(\t)
 \ee

 The eigenfunctions satisfy the following orthogonality and completeness conditions.

 \be\label{orthocond}
 \int_{-\infty}^{\infty} \Phi_l^{\dagger}(\t) \Phi_{l'}(\t) d\t =\delta_{ll'}~,~~~~\sum_l \Phi_l(\t)\Phi_l^{\dagger}(\t')=  \delta(\t-\t')\mathbb{I}
 \ee
 Beside these standard conditions the eigenfunctions must also satisfy certain continuity conditions. In view of the fact that the jump in the potential matrix is finite, the eigenfunctions and their first derivative must be continuous at $\t=0$. These conditions  will be translated into continuity conditions which we must impose on the associated heat kernel and the Green's function ( its Laplace Transform), and they will play a crucial role in constructing the  explicit form of the Green's function.

 Using  (\ref{heategf}) and (\ref{orthocond}) we can write (\ref{id}) as

 \be\label{heat1}
 \ln \mathrm{Tr}\rho_A^n=\frac{1}{2}\int_0^\infty \frac{1}{s}(\mathrm{Tr} \mathbb{ K}_n
 (\t,\t,s)-n\mathrm{Tr} K(\t,\t,s)) ds, ~~~K(\t,\t,s)= \mathbb{ K}_1(\t,\t,s)
 \ee

 Define now the Laplace transform of the heat kernel $\mathbb{ K}_n$ , $ \mathbb{ G}_n(\t,\t', E)= \int_{0}^{\infty} e^{-Es} \mathbb{ K}(\t,\t', s) ds$.  The $Nn\times Nn$ matrix function $\mathbb{ G}_n(\t,\t', E)$ is just the Green's function associated with  $-\frac{d^2}{d\t^2}+\mathbb{V}(\t) +E$  satisfying the following differential equation

 \begin{equation}\label{GE}
 (-\frac{d^2}{d\t^2}+\mathbb{V}(\t)+E)\mathbb{G}_n(\t,\t')=\delta(\t-\t') \mathbb{I},~~~~~G(\t,\t')=
 \mathbb{G}_1(\t,\t')
 \end{equation}

 it is then straightforward to show
 \be\label{Gh}
 \ln \mathrm{Tr}\rho_A^n=\frac{1}{2}\int_{0}^{\infty} (\mathrm{Tr} \mathbb{ G}_n(\t,\t,E)-n\mathrm{Tr} G(\t,\t,E)) dE
 \ee
 The trace $\mathrm{Tr}$ is understood to include integration over $\t$\footnote{Equation (\ref{Gh}) is indeed equivalent  to the evaluation of  $\ln \mathrm{Tr}\rho_A^n$ for massive scalar theory by first evaluating the derivative of  $\ln  \mathbb{Z}_n$   with respect to $m^2$, with the role of $m^2$ is here played  by $E$,  expressing it in terms of the coincident Green's function $\mathbb{ G}_n(\t,\t,m^2)$  on the $n$-fold cover space and then integrating with respect to $m^2$   \cite{Calabrese:2004eu}}.

 Let us now see how  $\mathbb{G}_n(\t,\t')$ can be constructed. In view of the fact that $ \mathbb{V}$
 and $\mathbb{V}_p$ do not generally commute, the construction of the Green's matrix is not straightforward.  The construction given here is actually slightly different from the original line of thought that led us to the correct form of this Green's matrix function, but it is more transparent and pedagogical.

 To construct $\mathbb{G}_n(\t,\t')$ we first start by considering the special case where $\mathbb{V}=V$ and $\mathbb{V}_p=V_p$ are just real scalars .  In this case   the problem is reduced to finding   the Green's function   for one-dimensional step potential for  Schrödinger   equation (with an appropriate redefinitions of the mass and other parameters). The Green's function for one dimensional step potential has been derived using different techniques  \cite{PhysRevLett.71.1, Acila_2006}. Using those  results the Green's function is given in the case   $\mathbb{V}=V$ and $\mathbb{V}_p=V_p$  by,

 \be\label{GF1d}
 G_{\mp\mp}(\t,\t') = \frac{1}{2{W}_\mp} e^{-{W}_\mp|\t-\t'|}
 +\frac{1}{2{W}_\mp}   \bigg[
 \frac{{W}_\mp-{W}_\pm}{{W}_-+{W}_+}  \bigg] e^{\pm{W}_\mp(\t+ \t')}
 \ee
 and

 \be\label{GF1d2}
 {G}_{\pm\mp}(\t,\t')= \frac{1}{{W}_-+{W}_+ }
 e^{\mp{W}_\pm\t \pm{W}_\mp \t'}
 \ee

 where we used the standard notation ${G}_{+-}(\t,\t')={G}_n(\t,\t') $ for $\t >0, \t'<0$
 ..etc and with
 $$
 {W}_{\pm}= \sqrt{{V}_\pm},~~{V}_- = {V}+E ,~~~~~{V}_+ {=V}_p
 +E
 $$

 Consider now the case where $\mathbb{V}$ and $\mathbb{V}_p$  are two commuting matrices. In this case the problem is practically equivalent to the one-dimensional case, and the Green's function keeps formally the same form without any ambiguity. Of course $\frac{1}{\mathbb{W}}$ and $\frac{1}{\mathbb{W}_-+\mathbb{W}_+ }$ would stand for the inverse of the matrices $\mathbb{W}$, and ${\mathbb{W}}_-+{\mathbb{W}}_+$, however we do not need to worry about the ordering of the different matrices appearing in  equations (\ref{GF1d}) and (\ref{GF1d2}).

 For the general case where $\mathbb{V}$ and $\mathbb{V}_p$ are non-commuting, equations (\ref{GF1d}) and (\ref{GF1d2}) do not work as they stand and the problem seems more challenging.

 We note here that we  are facing two  possibilities. Either we have just to arrange the matrices so that the resulting Green's matrix is still a solution and satisfies all the continuity and boundary conditions, or consider the possibility that   extra terms are present and which vanish in the commutative case, or  possibly considering more complicated form that involves chronological ordering operations. Fortunately it turns out that all we need is to use the Green's matrix for  commutative case and rearrange the matrices multiplications to obtain the correct solution.

 We first  note that although we were unable to obtain $\mathbb{G}(\t,\t')$  by directly solving the differential equation (\ref{GE}), it is possible to write down a series expansion for $\mathbb{G}(\t,\t')$ around the free one $\mathbb{V}(\t)=0$ . This is achieved  by writing  (\ref{GE}) as an integral equation  and using Laplace transform method to turn it into Riemann-Hilbert problem. Using the  technique known for these kind of problems one can show that $\mathbb{G}(\t,\t')$ has the following expansion up to the second order in $\mathbb{V}(\t)$ , for $\t,\t'<0$,

 \bea\label{GFapp}	
 \mathbb{G}_{--}(\t,\t')&=& \frac{e^{-\sqrt{E}|\t-\t'|}}{2\sqrt{E}} \bigg[ 1-\frac{\mathbb{V}}{2E}(\sqrt{E}|\t-\t'|+1) +\frac{1}{16}(\frac{\mathbb{V}}{E})^2\big[ (\sqrt{E}|\t-\t'|+3/2)^2 \nonumber \\
  &+&3/4\big] +..\bigg]
 +	\frac{e^{\sqrt{E}(\t+\t')}}{2\sqrt{E}}\bigg[ -\frac{\delta \mathbb{V}}{4 E} +\frac{1}{8}(\frac{\delta \mathbb{V}}{E})^2-\frac{\mathbb{V}\delta \mathbb{V}}{16E^2} (\sqrt{E}\t-3/2)\nonumber \\
 &-&\frac{\delta \mathbb{V}\mathbb{ V}}{16E^2} (\sqrt{E}\t'-3/2)+..\bigg]
 \eea

 where $\delta\mathbb{V}= \mathbb{V}-\mathbb{V}_p$.

 Similar expansion could be obtained for $G_{\pm\mp}$.

 The detail of   the derivation of the above expansion is given in Appendix I.

 Now it is not difficult to show that the expansion (\ref{GFapp}) and those for other regions  coincide with the expansion of the following expression,

 \be\label{GFS}
 \mathbb{G}_{\mp\mp}(\t,\t') = \frac{1}{2\mathbb{W}_\mp} e^{-\mathbb{W}_\mp|\t-\t'|}
 +\frac{1}{2\mathbb{W}_\mp}  e^{\pm\mathbb{W}_\mp\t}
 (\mathbb{W}_\mp-\mathbb{W}_\pm)(\mathbb{W}_-+\mathbb{W}_+)^{-1}   e^{\pm\mathbb{W}_\mp \t'}
 \ee

 and

 \be\label{GFS2}
 \mathbb{G}_{\pm\mp}(\t,\t')=  e^{\mp\mathbb{W}_\pm\t}\frac{1}{\mathbb{W}_-+\mathbb{W}_+ }
 e^{\pm\mathbb{W}_\mp \t'}
 \ee

 $$
 \frac{1}{\mathbb{W}_{\pm}}=\mathbb{W}_{\pm}^{-1},~~ \mathrm{and} ~~
 \frac{1}{\mathbb{W}_-+\mathbb{W}_+} \equiv
 (\mathbb{W}_-+\mathbb{W}_+)^{-1}
 $$
 where $\mathbb{W}_{\pm}$ are given by
 
 $$
 \mathbb{W}_-=   \mathbb{I}_{n} \otimes \sqrt{(V+E)} ,~~~~\text{and}~~ \mathbb{W}_+=\mathbb{P}_\pi \mathbb{W}_- \mathbb{P}_\pi^T
 $$

 The above matching between the expansion and the forms given by equations
 (\ref{GFS}) and (\ref{GFS2}) does not consist a full proof, nevertheless it  does almost bring  us home. Actually we can forget about the root of (\ref{GFS}) and (\ref{GFS2})  and limit our effort to showing  that they consist the correct solution  of (\ref{GE}).

 Let us remember that the Green function we are seeking for is uniquely defined   by the
 following  requirements:
 \begin{itemize}
 	\item  a- It satisfies the following equation
 	$$
 	(-\frac{d^2}{d\t^2}+\mathbb{V}(\t)+E)\mathbb{G}_n(\t,\t')=\delta(\t-\t'),~~~~~$$
 	\item b- $\lim_{\t,\t' \rightarrow \pm \infty} \mathbb{G}_n(\t,\t') =0$
 	\item c- All continuity conditions are satisfied.
 \end{itemize}

 Requirements (a) and (b) are obviously satisfied.  The continuity conditions are not obvious and needed to be listed and checked one by one.

 These conditions follow from the continuity conditions that the eigenfunctions and their first derivative must satisfy in view of the fact that the potential has a finite jump at the cut $\t =0$.

 \be\label{cc}
 \lim_{\t \rightarrow 0^-}\Phi_l(\t)=  \lim_{\t \rightarrow 0^+}\Phi_l(\t), ~~~\text{and} ~~\lim_{\t \rightarrow 0^-}\frac{d\Phi_l(\t)}{d \t}=  \lim_{\t \rightarrow 0^+}\frac{d\Phi_l(\t)}{d\t}
 \ee

 In Appendix I we list the resulting continuity conditions and show that they are all fulfilled .

 It should be noted here that beside these standard  continuity conditions there is an extra condition that the solution must satisfy, namely $ \mathbb{G}(\t,\t')^{\dagger}= \mathbb{G}(\t',\t)$ .  It follows from (\ref{heategf})  and we shall refer to it as the hermiticity constraint . In Appendix I this condition is shown to be fulfilled by $  \mathbb{G}(\t,\t')$ of (\ref{GFS}) and (\ref{GFS2}).  The hermiticity constraint  reduces by half the number of continuity conditions we must
 check, we will have to only consider the continuity with respect to one argument  of $\mathbb{G}(\t,\t')$.

 This completes the construction of the Green's function.

 Substituting now  $\mathbb{G}_n(\t,\t)$ into eqt (\ref{Gh}) and performing the integration with respect to $\t$ we obtain

 \be\label{ent3}
 \ln \mathrm{Tr}\rho_A^n=  \frac{1}{8}\int_{0}^{\infty} \mathrm{Tr}  (
 \mathbb{V}_-^{-1}-\mathbb{V}_+^{-1})(\mathbb{W}_--\mathbb{W}_+)\mathbb{(W}_-+\mathbb{W}_+)^{-1} dE
 \ee

 This  our alternative formula  which  gives R\'{e}nyi entropy  resulting from tracing out an arbitrary subset of H.Os, and it can be used  to compute the
 entanglement entropy via the replica trick.

 Some remarks about formula (\ref{ent3}) are in order.

 The expression of $\ln \mathrm{Tr}\rho_A^n$  is invariant under  rescaling of the potential and vanishes when $ [ \mathbb{V}, \mathbb{P}_\pi]=0$ as it should do. Actually it is the
 nonvanishing of the commutator $[ \mathbb{V}, \mathbb{P}_\pi] $  which will
 lead to non zero entanglement entropy, therefore $ \|[ \mathbb{V}, \mathbb{P}_\pi] \|/\|\mathbb{V}\| $ can be considered in a
 sense a measure of the degree of entanglement. The matrix norm $\|.\|$ could be the Hilbert-Schmidt norm or whatever matrix norm one prefers.  Actually it turns out that when using the weak norm it is more   appropriate to rescale the commutator and use   instead  $\sqrt{N} \|[ \mathbb{V}, \mathbb{P}_\pi] \|/\|\mathbb{V}\| $ as a measure of the degree of entanglement.

 Finally, we note that in principle one can develop a similar formalism for a Fermionic lattice system by using Euclidean path integral based on Grassman variables, imposing  anti-periodic conditions and derive the corresponding Green's matrix and EE formula . This would offer a new tool to investigate the EE for several Fermionic models in connection with critical phenomena and RG, like the Long-Range Kitaev chain and other models.
\section{Massive Scalar Theory on 1-d lattice}

 As  a first application and good warm-up exercise we apply (\ref{ent3}) to free massive K.G field on  1+1-d Minkowski spacetime. This model in the continuum has been investigated using different techniques.   As a by-product of this problem    we shall rederive the standard formula of the EE for  two coupled H.O's.

 The Lagrangian is given by

 \be\label{scalarlag}
 L=\frac{1}{2}\int dx \big((\partial _\mu \phi)^2 +m^2 \phi^2\big)
 \ee

 We put this model on a lattice as follow .
 The $x-$axis is replaced by a one-dimensional lattice,
 i.e $ x_i = i\epsilon$ where $\epsilon$ is the lattice spacing. As an infrared cuttoff we take the size of the lattice to be $R=2N\epsilon$ and impose periodic boundary conditions .

 $$
 \phi(x) \rightarrow q_i,~~~~~~i=-(N-1).....(N-1)
 $$

 By rescaling the field variable we obtain a dimensionless Lagrangian

 \be
 L_0=\frac{1}{2}( \dot{Q}^T \dot{Q} +Q^T V Q)
 \ee

 The potential matrix $V$ is $2N\times 2N $ matrix  given by

 \be\label{p}
 V=
 \left(
 \begin{array}{cccccccc}
 	2+\mu^2 & -1 & 0 & 0& 0&  \dots  & 0& -1 \\
 	-1 & 2+\mu^2 &-1 &0 & 0 & \dots  & 0& 0 \\
 	0 &-1 &2+\mu^2 & -1 & 0 & \dots & 0 & 0\\
 	0 & 0 &-1 & 2+\mu^2 & -1 & \dots & 0& 0\\
 	0 & 0 & 0 & -1 & 2+\mu^2 & \dots & 0& 0\\
 	\vdots & \vdots & \vdots & \vdots & \ddots & \vdots& 2+\mu^2& -1 \\
 	-1& 0 & 0 & 0 & \dots & 0 &-1& 2+\mu^2
 \end{array}
 \right)
 \ee
 Where $\mu=m \epsilon$.

 We note that $V$ is a circulant symmetric real matrix  and therefore its elements can be written using its eigenvalues as

 \be\label{potential1}
 V_{lm}=\frac{1}{2N}\sum_{k=0}^{2N-1} (\mu^2 +2(1-\cos \pi k/N))e^{i\pi k (l-m)/N}
 \ee

 \subsection{The Half-Space Entanglement }

 Let us consider the entanglement entropy resulting from tracing out the positive axis or the set $A=(q_N,q_{N+1}, \cdots q_{2N-1})$.

 We need first to compute $(\mathbb{V}_-^{-1}-\mathbb{V}_+^{-1})$ , $(\mathbb{W}_{+}-\mathbb{W}_{-})$ and $(\mathbb{W}_{+}+\mathbb{W}_{-})$ .

 For $\mathbb{V}_-$ we have $ \mathbb{V}_-=   \mathbb{I}_{n} \otimes (V+E)$, and $\mathbb{V}_-^{-1}= \mathbb{I}_{n} \otimes (V+E)^{-1}$. $ \mathbb{V}_+^{-1}$ is obtained by applying $\mathbb{P}_\pi(A)$ , $\mathbb{V}_+^{-1}=\mathbb{P}_\pi \mathbb{V}_-^{-1} \mathbb{P}_\pi^T$ .
 Similarly for $\mathbb{W}_-=   \mathbb{I}_{n} \otimes \sqrt{(V+E)} $ , and $\mathbb{W}_+=\mathbb{P}_\pi \mathbb{W}_- \mathbb{P}_\pi^T$.

 In view of the fact that  $V+E$ is circulant symmetric matrix, both   $(V+E)^{-1}$ and $\sqrt{(V+E)}$ are also  circulant symmetric matrices which can be decomposed as
 \be
 (V+E)^{-1}=
 \left(
 \begin{array}{cc}
 	A & B \\
 	B^T &A \\
 \end{array}
 \right)
 ~~~~,~~\sqrt{(V+E)}=\left(
 \begin{array}{cc}
 	G & F \\
 	F^T &G \\
 \end{array}
 \right)
 \ee

 $A, B, G,F$ are $N \times N$ matrices.

 Now, putting everything together and using (\ref{matrix}) it is straightforward to show  that

 \be
 (\mathbb{V}_-^{-1}-\mathbb{V}_+^{-1})(\mathbb{W}_{+}-\mathbb{W}_{-})=  \mathbb{C} \otimes  B.F
 \ee

 where $ \mathbb{C}$ is a $2n\times 2n$ circulant matrix with elements
 $ c_j$, $j=0,1.\cdot \cdot\cdot 2n-1$, $c_0=2, c_{2n-2}=c_2=-1$  and zeros otherwise.

 And

 \be
 (\mathbb{W}_{+}+\mathbb{W}_{-}) =
 \left(
 \begin{array}{cccccccc}
 	2G & F & 0 & 0& 0&  \dots  & 0& F\\
 	F& 2G &F &0 & 0 & \dots  & 0& 0 \\
 	0 &F &2G & F & 0 & \dots & 0 & 0\\
 	0 & 0 &F & 2G &F & \dots & 0& 0\\
 	0 & 0 & 0 & F & 2G & \ddots & 0& 0\\
 	\vdots & \ddots & \ddots & \ddots & \ddots & \ddots& \ddots&\vdots \\
 	\vdots & \ddots & \ddots & \ddots & \ddots & \ddots& 2G& F \\
 	F& 0 & 0 & 0 & \dots & 0 &F& 2G
 \end{array}
 \right)
 \ee

 Therefore we can use  discrete Fourier transformation (DFT) to simultaneously block diagonalize  $ (\mathbb{W}_{+}+\mathbb{W}_{-})$ and $(\mathbb{V}_-^{-1}-\mathbb{V}_+^{-1})(\mathbb{W}_{+}-\mathbb{W}_{-})$ ,  perform partial inversion of $(\mathbb{W}_{+}+\mathbb{W}_{-})$   and get

 \bea\label{tr1}
 \mathrm{Tr}  (
 \mathbb{V}_-^{-1}-\mathbb{V}_+^{-1})(\mathbb{W}_--\mathbb{W}_+)\mathbb{(W}_-+\mathbb{W}_+)^{-1}=\sum_{j=0}^{2n-1}(1-\cos\frac{2\pi j}{n})\mathrm{Tr} B.F T^{-1}
 \eea
 where
 $$
T=G+F\cos\frac{\pi j}{n}
$$
 and we obtain for  R\'{e}nyi entropy

 \be\label{renyi1}
 \ln\mathrm{Tr}\rho_A^n=-\frac{1}{8}\sum_{j=0}^{2n-1}(1-\cos\frac{2\pi j}{n}) \int_{0}^{\infty}\mathrm{Tr_N} B.F (G+F\cos\frac{\pi j}{n})^{-1} dE	
 \ee

 The matrices elements of $B,F $ and $G$  follow from (\ref{potential1}),

 \be\label{1}
 B_{ml}=\frac{1}{2N} \sum_{k=0}^{2N-1} \varphi_k^{-1} (-1)^k e^{i\pi k(m-l)/N}
 \ee
 \be\label{2}
 F_{ml}=\frac{1}{2N} \sum_{k=0}^{2N-1} \varphi_k^{1/2} (-1)^k e^{i\pi k(m-l)/N}
 \ee
 \be\label{3}
 G_{ml}=\frac{1}{2N} \sum_{k=0
 }^{2N-1} \varphi_k^{1/2}  e^{i\pi k(m-l)/N}
 \ee

 where $\varphi_k= \mu^2+E+2(1-\cos\pi k/N)$, and $l,m$ run now over the rang $ l,m=0,\dots N-1$.

 Before we move on in studying the large $N$ and small $\mu$ behavior of R\'{e}nyi entropy for this model let us first use (\ref{renyi1}) in the special  case $N=1$ to obtain the standard EE formula for two coupled H.O's.

 In the case $N=1$ the potential (\ref{potential1}) becomes

 \be\label{potentialHO}
 V=
 \left(
 \begin{array}{cc}
 	2+\mu^2+E &-2 \\
 	-2 &2+\mu^2 +E\\
 \end{array}
 \right)
 \ee

 which corresponds to a rescaled potential describing two coupled H.O's with equal frequencies. The case of two coupled H.O's with different frequencies is a little bit more complicated technically and can not be deduced from this model, but it can be worked out exactly \cite{AlloucheThesis}.

 Now, substituting  $N=1$ in (\ref{renyi1})	and using  (\ref{1}), (\ref{2}) and (\ref{3}) we obtain

 \be
 \ln \mathrm{Tr}\rho_A^n= -\frac{1}{8}\sum_{j=0}^{2n-1} (1-\cos \frac{2\pi j}{n})\int_{a_0}^{\infty} \frac{\sqrt{x-1}-\sqrt{x+1}}{(x^2-1)(\a_+\sqrt{x-1}+\a_-\sqrt{x+1})} dx
 \ee

 $\a_{\pm}=1\pm\cos\frac{\pi j}{n}$, $a_0 =\mu^2/2+1$.

 The integral inside the sum can be exactly evaluated and after some arrangements and  simplifications we end up with

 \be\label{sum}
 \ln \mathrm{Tr}\rho_A^n= -\sum_{j=1}^{n-1} \ln  \frac{\cos \frac{\pi j}{n}+\b }{\sqrt{\b^2-1}}
 \ee

 where $\b=a_0+\sqrt{a_0^2-1}$.

 The sum  in (\ref{sum} ) gives
 \be\label{reyni}
 \ln \mathrm{Tr}\rho_A^n= \ln \frac{(\a^2-1)^{n}}{(\a^{2n}-1)}, ~~~~~\a=\b+\sqrt{\b^2-1}
 \ee

 Here our $\alpha$ turns out to be related to $\xi$ defined in \cite{Srednicki:1993im} by $\a=\frac{1}{\sqrt{\xi}}$.

 Using the replick trick and the definition of $\a$ we recover the standard formula for EE for two coupled H.O's,
 \be\label{reyni2}
 -\frac{d}{dn}\ln \mathrm{Tr}\rho_A^n|_{n=1}= -\ln (1-\xi) -\frac{\xi}{1-\xi}\ln \xi
 \ee

 Having  derived the standard formula for EE for 2-coupled H.O let us now focus on the large $N$ limit .

 The main technical obstacle is  the  inverse of the matrix $G+F\cos\frac{\pi j}{n}$, therefore we shall dedicate a special subsection to this problem .

 We first note that our goal is  not  to find an exact inverse to this matrix for arbitrary $N$, but we  are only interested in the large $N$ limit or an asymptotic inverse.

 We start by rewriting  equation ( \ref{renyi1}) as

 \be\label{renyimod}
 \ln\mathrm{Tr}\rho_A^n=-\frac{1}{16}\sum_{j=0}^{2n-1}(1-\cos\frac{2\pi j}{n}) \int_{a_0}^{\infty}\mathrm{Tr} \tilde{B}.\tilde{F} T_N(\alpha)^{-1} da	
 \ee

 where $T_N(\alpha)$ is given in the large $N$ limite by
 \be\label{Toeplitz}
 T_N(\alpha)=  \alpha_+ C+\alpha_-\tilde{C}
 \ee

 \be\label{Cira}
 C_l=\frac{1}{N} \sum_{k=0}^{N-1} \sqrt{c_{k}} e^{i2\pi k l/N},~~\tilde{C}_l=\frac{e^{i\pi l/N}}{N} \sum_{k=0}^{N-1}\sqrt{c_{k}} e^{i2\pi k l/N}
 \ee

 where  $\alpha_{\pm} = 1\pm \cos\frac{\pi j}{n} $, and  $ c_k =a-\cos \frac{2\pi k}{N} $.

 \be\label{matriceBF}
 \tilde{B}_{l}=\frac{1}{N} \sum_{k=0}^{2N-1} \tilde{\varphi}_k^{-1} (-1)^k e^{i\pi kl/N}, ~~
 \tilde{F}_{l}=\frac{1}{N} \sum_{k=0}^{2N-1} \tilde{\varphi}_k^{1/2} (-1)^k e^{i\pi kl/N}
 \ee

 \be
 \tilde{\varphi}_k = a-\cos \frac{\pi k}{N}, ~~ a= \frac{\mu^2+E}{2}+1
 \ee

 \subsection{On the Asymptotic Inverse of a Particular Class of Toeplitz Matrices}

 The matrix $T_N(\a)$ defined by (\ref{Toeplitz}) and (\ref{Cira}) is a particular Toeplitz matrix, a linear combination of a circulant and anti-circulant matrices.  By anti-circulant matrix we refer to the matrix $\tilde{C}_l$. Both $C$ and $\tilde{C}$ can be inverted  separatly, however their linear combination has no known inverse. Moreover in the large $N$ limit $ T_N(\a)$ does not admit a Wiener representation with $N$-independent symbol. Although a great deal is known about the asymptotic behavior of Toeplitz matrix with $N$-independent symbol under certain conditions, very little, if any, is known about the asymptotic behavior of Toeplitz matrices with $N$-dependent symbol, as it is the case for $T_N(\a)$. This kind of matrices is expected to show up within this approach whenever we trace out half of the space, therefore it is necessary to understand the the large $N$ or asymptotic behavior of these type of matrices and their inverses.

 Before tackling the problem of finding an asymptotic inverse for $T_N(\a)$, let us see what the available mathematical literature can say about this inverse.

 It can be shown using the machinery of Generalized Locally Toeplitz  (GLT)  \cite{https://doi.org/10.1002/nla.2286, Barbarino} that the trace of $T_N(\a)^{-1}$ has the following asymptotic limit\footnote{ We are indebted to A. Böttcher,  M.Halwass and S.Serra-Capizzano for this result.}

 \be\label{traceinvese}
 \lim_{N\rightarrow \infty} \frac{ \mathrm{Tr} T_N(\alpha)^{-1}}{N}=\frac{1}{\pi}\int_{0}^{\pi} \frac{d\theta}{\sqrt{a-\cos \theta}}=\frac{1}{\pi\sqrt{a+1}} K(\sqrt{\frac{2}{a+1}})
 \ee

 where $K$ is the complete elliptic integral of the first  kind.

 The above result can be easily be confirmed numerically. At $N=50$ and for several values of $a$ and $n$ we get a perfect coincidence.

 We shall not enter into the detail of the derivation of the above asymptotic trace of the inverse, because as it unfortunately turned out the GLT machinary and their $*$-algebra structure fails to give a good estimation of the full trace which appears in eqt (\ref{renyimod}),  they only give weak information on this trace not strong enough to be of any use for our particular problem, namely being just of order one.  Nevertheless we shall use the above result about the asymptotic behavior of the trace  as a guide in our search for the inverse .

 We begin our search for the inverse by noting that it must be a persymmetric matrix as can easily be shown. Analytical and numerical investigation of the structure of both $T_N({\a})$ and its inverse led us to consider  the following ansatz as a candidate for the inverse.

 \be\label{ansatz}
 (M_N)_{lm}= \frac{1}{N^2}\sum_{p,k=0}^{N-1} \big[\alpha_{kp}e^{-i\pi m/N}k_+(p-k)+\alpha_{pk} e^{i\pi l/N} k_-(p-k) \big]e^{i 2\pi kl/N} e^{-i2\pi pm/N}
 \ee

 where $ \a_{pk}$ is a function of $p$ and $ k$ to be determined later, and $ k_{\pm}$ are two  functions ( a sort of kernels) defined by

 \be
 k_{\pm}(x)= \frac{1}{2}\sum_{m=0}^{N-1} e^{\frac{2\pi i}{N} (x\pm1/2)m}=\frac{1}{1-e^{2\pi i (x\pm1/2)/N}}
 \ee

 It is easy to see that $ M_N$ is persymmetric matrix for arbitrary  $\a_{pk}$.

 Now, the ideal thing is to fix $\a_{pk}$ by solving $ M_N T_N=I$, for finite or atleast large $N$. Unfortunately this direct approach led to apparently unsolvable equations. However it turns out that a particular value for $\a_{pk}$  solves the following   necessary condition $\mathrm{Tr} M_N T_N =N  $.

 Using the form of $T_N$ and the ansatz $M_N$ it is a straightforward exercise to show that with $\a_{pk}$ given by

 \be\label{alphapk}
 \alpha_{kp}=  \frac{1}{\alpha_+\sqrt{c_k}+\alpha_-\sqrt{c_p}} \Rightarrow  \mathrm{Tr} M_N T_N =N
 \ee	

 Although this particular choice of $\a_{pk}$ only solves the trace condition ( for arbitrary $N$!) it turns out to be a good candidate for an asymptotic inverse for $T_N$.

 Before discussing the validity of $M_N$, with $\a_{pk}=\frac{1}{\alpha_+\sqrt{c_k}+\alpha_-\sqrt{c_p}}$,
 as an asymtotic inverse, some few remarks about this ansatz are in order\footnote{From now on $M_N$ will refer  to the original ansatz with  $\a_{pk}=\frac{1}{\alpha_+\sqrt{c_k}+\alpha_-\sqrt{c_p}}$.}

 It is easy to show that  $M_N$ gives the exact inverse for the two particular cases  $( \alpha_-=0, \alpha_+\ne 0)$ and $( \alpha_+=0, \alpha_+\ne 0)$.  For large $a$, the expansion  of $ M_N$  coincides with the neuman series of the inverse of $T_N$, at least for the first few orders we checked explicitly.

 Now, what about  $a$ close to one, the case of most interest to us?.

 We have run serval numerical calculations for the following norms (the weak norm or Hilbert-Shmidth).

 $$
 \delta_N(a)= | T_N.M_N-I|
 $$
 and

 $$
 \Delta_N(a) = |M_N-T_N^{-1}|
 $$

 For both $\delta_N(a)$ and $\Delta_N(a)$ and as long as $a$ is not very close to $1$, $a=1.2$ and larger, they are both smaller than $10^{-3}$ for $N=30$, and becoming really negigeable as we reach $1.5$, larger than this there is almost a perfect matching (for $a=2$ both $\delta$ and $\Delta$ are of the order $10^{-4}$ ). Of course they become smaller as we move to larger $N$.

 For $a$ close to one, $ a=1.001$ say, one needs to go to larger $N$ to get small values for $\delta$ and $\Delta$, however even for $N=40$ and $N=50$ the two norms are of the order of $1/N$ numerically. For $a=1.002, N=100$ , $\Delta=0.06$. These numerical results provide strong evidences that both  $\delta_N(a)$ and $\Delta_N(a)$ are of the order of $1/N$ or atleast vanish in the large $N$ limit, as long as $a> 1$. Therefore we shall make the following  conjecture.

 \textbf{\emph{Conjecture}}: Let $T_N$ be a Toeplitz matrix given by

 $$
 T_N(\alpha)=  \alpha_+ C+\alpha_-\tilde{C}
 $$

 $$
 C_l=\frac{1}{N} \sum_{k=0}^{N-1} \sqrt{c_{k}}e^{i2\pi k l/N},~~\tilde{C}_l=\frac{e^{i\pi l/N}}{N} \sum_{k=0}^{N-1}\sqrt{c_{k}} e^{i2\pi k l/N}
 $$

 where  $\alpha_{\pm} = 1\pm \cos\frac{\pi j}{n} $, and
 $ c_k =a-\cos \frac{2\pi k}{N} $.

 then the following matrix

 $$
 (M_N)_{lm}= \frac{1}{N^2}\sum_{p,k=0}^{N-1} \big[\alpha_{kp}e^{-i\pi m/N}k_+(p-k)+\alpha_{pk} e^{i\pi l/N} k_-(p-k) \big]e^{i 2\pi kl/N} e^{-i2\pi pm/N}
 $$

 with
 $$
 k_{\pm}(x)=\frac{1}{2} \sum_{m=0}^{N-1} e^{\frac{2\pi i}{N} (x\pm1/2)m},~~~~ \a_{pk}= \frac{1}{\alpha_+\sqrt{c_k}+\alpha_-\sqrt{c_p}}
 $$

 is an asymptotic inverse of $T_N(\a)$.

 By asymptotic inverse here we mean that $M_N$ fulfil the following condition
 $$
 \lim_{N\rightarrow\infty}  \delta_N(a)= | T_N.M_N-I| \rightarrow 0, ~~\forall a > 1
 $$

 We shall not try to prove this conjecture at this stage nor do we  pretend to know how  $\delta_N(a)$ approaches zero, although our numerical calculation suggests it vanishes  like $1/N$.

 In principle  $\delta_N(a)$ can be computed explicitly, we have the analytical expression of both $T_N$ and $M_N$,  however upon first inspection it seems to be too complicated to handle directly analytically. On the other hand and interestingly   our numerical studies of several Toeplitz matrices similar to $T_N$ but with different $c_k$ and $\a_{\pm}$, showed that our ansatz extends beyond our particular problem. More precisely we have.

 If $ T_N$ is a Toeplitz matrix given by

 $$
 T_N(\alpha)=  \alpha C_1+\beta\tilde{C_1}
 $$

 $$
 C_l=\frac{1}{N} \sum_{k=0}^{N-1} a_k e^{i2\pi k l/N},~~\tilde{C}_l=\frac{e^{i\pi l/N}}{N} \sum_{k=0}^{N-1}a_{k} e^{i2\pi k l/N}
 $$

 $a_k$ and $b_k$ are two  functions of $k$, which need to satisfy certain technical conditions otherwise arbitrary.

 Then the following matrix

 $$
 (M_N)_{lm}= \frac{1}{N^2}\sum_{p,k=0}^{N-1} \big[\alpha_{kp}e^{-i\pi m/N}k_+(p-k)+\alpha_{pk} e^{i\pi l/N} k_-(p-k) \big]e^{i 2\pi kl/N} e^{-i2\pi pm/N}
 $$

 with
 $$
 \a_{pk}= \frac{1}{\alpha{a_k}+\beta{b_p}}
 $$

 is an asymptotic inverse for $T_N$ with $\lim_{N\rightarrow \infty}\delta_N \rightarrow 0$.

 This results, if confirmed by analytic calculation, will offer a general solutions for the asymptotic inverse of a wide class of Toeplitz matrices, with $N$-dependent symbol, which are not covered by the known results in literature \cite{grenandertoeplitz, widom1974asymptotic}. Therefore a more mathematically oriented work is needed to settle this issue.

 Now, let us  test our conjecture by comparing   the trace of $M_N$ for $N\rightarrow \infty$   with the asymptotic result obtained  using GLT machinery, namely equation (\ref{traceinvese}).

 Using equation (\ref{ansatz}) is straightforward to show

 \be\label{traceans1}
 \frac{\mathrm{Tr}M_N(\alpha)^{-1}}{N}=\frac{1}{N^2}\sum_{l=-(N-1)}^{N-1} (1-\frac{|l|}{N}) \alpha_{kp} e^{i2\pi(p-k+1/2)l/N}
 \ee

 In the limit $N \rightarrow \infty$ we can turn the above  sums over $p$ and $k$ into  complex integrations around the unit circle and show

 \be\label{traceans}
 \lim_{N\rightarrow \infty}\frac{\mathrm{Tr}M_N(\alpha)^{-1}}{N}=\frac{1}{\pi\sqrt{a+1}} K(\sqrt{\frac{2}{a+1}})
 \ee
This result is proved in Appendix II.

The result we have obtained using our ansatz is identical to the asymptotic limit obtained using GLT machinery, equation (\ref{traceinvese}).

 Now, although the above result for the trace does not consist a proof for our conjecture it  however provides, along with the numerical results, strong supporting evidences for the its
 truthfulness. A complete proof of the above conjecture and its use to  extract the universal divergence for the Half-space EE in 1+1 will be the subject of a subsequent work.

 \subsection{Infinitismal Intrerval: One point Entanglement }

 One of the interesting cases which has been discussed extensively in literature is EE resulting from tracing out a single finite interval in 1+1. The universal part of the EE in 1+1 has been derived using different techniques, see \cite{Casini:2009sr,Calabrese:2004eu} and references therein.  This leading universal divergences is given for a massive scalar theory by

 \be\label{EE_interval}
 S=\frac{1}{3} \ln \frac{L}{\epsilon}+\dots
 \ee

 $L$ is the length of the interval and $\epsilon$ is the UV cutoff of the theroy.

 The above universal leading divergent term of EE is generally derived by assuming $L\gg \epsilon$. When the ignored region is of the order of the UV cutoff  the subsystem
 begins to fill most of the universe, thus there is less information to be lost by not
 measuring outside, and  eqt (\ref{EE_interval})  suggests that EE becomes
 of $O(1)$, which would be  dependent on the details of
 the regularization and looses physical meaning . However, with more elaborated tools it was subsequently possible  to show the existence of an extra IR divergence (softer)   in the massless limit, see  for instance \cite{Casini:2009sr} .

 More precisely we have for $m \rightarrow 0$ the following behavior for the entropy difference

 \be
 S(L)-S(L_0) \sim \frac{1}{3} \ln (L/L_0) +\frac{1}{2} \ln (-\ln (m L) )-\frac{1}{2} \ln (-\ln (m L_0) )
 \ee
 which suggests an IR divergence given by $	S(L)=\frac{1}{2} \ln (-\ln (m) )$ for any set. This IR divergence was given a simple heuristic explanation in terms of the zero mode of the quantum field in the massless limit. As a consequence the mutual information between two sets $A, B$ , $ I(A,B)\sim \frac{1}{2} \ln (-\ln (m) )$ is also IR divergent \cite{Casini:2009sr}.

 According to these results one would expect the entropy of an interval with size comparable  to the UV cutoff  to be also divergent with a universal coefficient  instead of being of the order of one as it was first anticipated, and this is what we are going to show using  the approach developed in this paper.

 For purely technical reasons we  shall  slightly modify our lattice  regularization and add one point in the middle of the lattice considered previously, the origin; to obtain  lattice with $2N+1$ point . We impose periodic boundary conditions, the  size of the lattice becomes $R=(2N+1)\epsilon $, with $2N+1$ point.
 $$
 \phi(x) \rightarrow q_i,~~~~~~i=-N..,0,..N
 $$

 Consider the entanglement entropy resulting from tracing out the set $A$ made of the point $q_0$.

 Writing $\mathrm{Tr}\rho_{q_0}$ using Euclidean path integral , equation (\ref{pi1}),   the  corresponding permutation matrix $\mathbb{P}_{\pi}$  can easily be shown to be given by

 \begin{eqnarray}
 \mathbb{P}_{\pi}(\{q_0\})=\left(
 \begin{array}{cccccc}
 {P}_{2N}& 0_{2N} & 0_{2N}&  \dots & 0_{2N}& P_0\\
 P_0& P_{2N}& 0_{2N}&  \dots  & 0_{2N}& 0_{2N}\\
 0_{2N}& P_0& P_{2N}& \dots & 0_{2N}& 0_{2N}\\
 \vdots& \vdots& \vdots&\ddots & \vdots& \vdots\\
 0_{2N}& 0_{2N}& \dots& P_0 & P_{2N}& 0_{2N}\\
 0_{2N}& 0_{2N}& 0_{2N}& \dots & P_0& P_{2N}
 \end{array}
 \right)
 \end{eqnarray}

 Where $0_{2N}$ stands for $(2N+1)\times (2N+1)$ zero matrix, $P_{2N}$ and $P_0$ are $(2N+1)\times(2N+1)$  projectors matrices on the available and unavailable regions defined as follows :
 \begin{eqnarray}
 P_{2N}=\left(
 \begin{array}{ccc}
 I_N& 0& 0_N\\
 0& 0& 0\\
 0_N& 0 & I_N
 \end{array}
 \right),\ \ P_0=\left(\begin{array}{ccc}
 0_N& 0& 0_N\\
 0& 1 & 0\\
 0_N& 0& 0_N
 \end{array}
 \right)
 \end{eqnarray}

 $I_N$ is the identity $N\times N$ matrix. $0_N$ is $N\times N$ zero matrix\footnote{ The other zeros are either row or column vectors  line vectors with zero entries, or just zero, accorrding to their position in the matrix}.  $P_0$ projects onto the subset $A={q_0}$ and $P_{2N}$ projects onto the complement made of the $2N$ points.

 The form of the permutation matrix suggests  the following decomposition of the potential matrix $V$
 \begin{eqnarray}
 V_-= V+E=\left(\begin{array}{ccc}
 A& b& B\\
 b^T& 2a& b'^T\\
 B^T& b'& A
 \end{array}
 \right)
 \end{eqnarray}

 Where $A$ and  $B$ are $N\times N$ matrices, $b$ and $b'$ are column
 vectors   and $a$ is the same parameter defined in the previous subsection,  $ a=\frac{\mu^2+E}{2}+1 $. A similar decompositions for $V_-^{-1} $ and $W_- $ follow
 \begin{eqnarray}
 V_-^{-1}=\left(
 \begin{array}{ccc}
 C& d& D\\
 d^T& c& d'^T\\
 D^T& d'& C
 \end{array}
 \right)
 ,\ \ \ W_-=
 \left(
 \begin{array}{ccc}
 G& f& F\\
 f^T& e& f'^T\\
 F^T& f'& G
 \end{array}
 \right)
 \end{eqnarray}

 The explicite forms of all these matrices and vectors follow from  eqt (\ref{potential1}) .

 The next step is to evaluate $\mathbb{V_+}^{-1}$ and $\mathbb{W_+}$.

 Using the explicite form of the permutation matrix it is easy to show that
 \begin{eqnarray}
 \mathbb{V}_+^{-1}=\mathbb{P}_{\pi}\mathbb{V}_-^{-1}\mathbb{P}^{T}_{\pi}&=& \left(
 \begin{array}{ccccccc}
 M_C& M_d& 0&0 &\dots& 0& M_d^T\\
 M_d^T& M_C& M_d& 0& \dots& 0& 0\\
 0& M_d^T& M_C& M_d& \dots& 0& 0\\
 \vdots& \vdots& \vdots& \vdots&\ddots& \vdots&\vdots\\
 0&  0& \dots& M_d^T & M_C& M_d& 0\\
 0&  0& 0& \dots& M_d^T& M_C& M_d\\
 M_d&  0& 0& \dots&0& M_d^T& M_C
 \end{array}
 \right)
 \end{eqnarray}

 \begin{eqnarray}
 \mathbb{W}_+=\mathbb{P}_{\pi}\mathbb{W}_-\mathbb{P}^{T}_{\pi}&=&
 \left(
 \begin{array}{ccccccc}
 M_G& M_f& 0& 0& \dots& 0& M_f^T\\
 M_f^T& M_G& M_f& 0& \dots& 0& 0\\
 0& M_f^T& M_G& M_f& \dots& 0& 0\\
 \vdots& \vdots& \vdots& \vdots&\ddots& \vdots&\vdots\\
 0&  0& \dots& M_f^T & M_G& M_f& 0\\
 0&  0& 0& \dots& M_f^T& M_G& M_f\\
 M_f&  0& 0& \dots&0& M_f^T& M_G
 \end{array}
 \right)
 \end{eqnarray}

 Where
 \begin{eqnarray}
 M_C=\left(	\begin{array}{ccc}
 C& 0& D\\
 0& c& 0\\
 D^T& 0& C
 \end{array}
 \right),\ \ M_d=\left(\begin{array}{ccc}
 0& d& 0\\
 0& 0&0\\
 0& d'& 0
 \end{array}
 \right),\ \ M_G=\left(\begin{array}{ccc}
 G& 0& F\\
 0& e& 0\\
 F^T& 0& G
 \end{array}
 \right),\ \
 \nonumber
  \end{eqnarray}

  \begin{eqnarray}
  M_f= \left(\begin{array}{ccc}
 0& f& 0\\
 0& 0& 0\\
 0& f'& 0
 \end{array}
 \right)\nonumber
 \end{eqnarray}

 Putting everything together  the following product can be readily computed
 \begin{eqnarray}
 (\mathbb{V}_-^{-1}-\mathbb{V}_+^{-1})(\mathbb{W}_--\mathbb{W}_+)=
 \left(	\begin{array}{ccccccc}
 2M& -M& 0& 0& \dots & 0& -M\\
 -M& 2M& -M& 0& \dots & 0& 0\\
 0& -M& 2M& -M& \dots & 0& 0\\
 \vdots& \vdots& \vdots& \vdots& \ddots & \vdots& \vdots\\
 0&  0& \dots&- M & 2M& -M& 0\\
 0&  0& 0& \dots&- M& 2M& -M\\
 -M& 0& 0& 0& \dots& -M& 2M
 \end{array}
 \right)
 \end{eqnarray}
 where $M=(M_d+M_d^T)(M_f+M_f^T)$.The resulting matrix is block circulant.

 On the other hand  the  matrix $\mathbb{W_-}+\mathbb{W}_+ $ is  block circulant as well :
 \begin{eqnarray}
 \mathbb{W}_-+\mathbb{W}_+= \left(\begin{array}{ccccccc}
 W_-+M_G& M_f^T& 0&0&\dots& 0& M_f\\
 M_f& W_-+M_G& M_f^T&0&\dots& 0& 0\\
 0& M_f& W_-+M_G& M_f^T& \dots & 0& 0\\
 \vdots& \vdots& \vdots& \ddots& \vdots& \vdots&\vdots\\
 0&  0& \dots& M_f & W_-+M_G& M_f^T& 0\\
 0&  0& 0& \dots& M_f& W_-+M_G& M_f^T\\
 M_f^T& 0& 0& \dots &0& M_f& W_-+M_G
 \end{array}
 \right)\nonumber
 \end{eqnarray}
 This allows us to block diagonalize the following  matrices product $(\mathbb{V}^{-1}_--\mathbb{V}^{-1}_+)(\mathbb{W}_--\mathbb{W}_+)(\mathbb{W}_- +\mathbb{W}_+)^{-1}$  using  DFT  and obtain the follwoing identity
 \begin{eqnarray}
 \mathrm{Tr}((\mathbb{V}_-^{-1}-\mathbb{V}^{-1}_+)(\mathbb{W}_--\mathbb{W}_+)(\mathbb{W}_-+\mathbb{W}_+)^{-1})=2\sum_{j=0}^{n-1}( 1-\cos(\frac{2\pi j}{n}) \mathrm{Tr} M T^{-1}
 \end{eqnarray}

where
$$
T=W+M_E+e^{-\frac{2\pi i j}{n}}M_f+e^{\frac{2\pi i j}{n}}M_f^T
$$

 Evaluating the above trace requires determining the inverse of the matrix $T$ given by the above equation.

 Note first that this matrix can be rewritten as
 \begin{eqnarray}
 T=	W+M_E+e^{-\frac{2\pi j}{n}}M_f^T+e^{\frac{2\pi i j}{n}}M_f=2W+\zeta^* M_f^T+\zeta M_f
 \end{eqnarray}
 Where $\zeta=1+e^{\frac{2\pi ij}{n}}$. Second  $W$ is circulant invertible matrix and both $M_f$ and $M_f^T$ are  of rank-one, therefore  by  recursive use of   Miller theorem \cite{Miller} we can explicitly invert $T$ :
 \begin{eqnarray}
 (2W+\zeta^*M_f)^{-1}_{lm}&=&\frac{W^{-1}_{lm}}{2}+\frac{\zeta^*}{2\Delta}(\delta_{Nm}-W_{NN}W^{-1}_{Nm})
 \end{eqnarray}

 \begin{eqnarray}
 T_{lm}^{-1}&=&	(2W+\zeta^*.M_f+\zeta.M_f^T)^{-1}_{lm}= (2W+\zeta^*M_f)^{-1}_{lm}+\frac{\zeta\Delta}{\Delta'}(\ \delta_{lN}-W_{NN}.W^{-1}_{lN}\nonumber\\
 &+&\frac{\zeta^*}{\Delta}W_{NN}W_{lN}^{-1}( W_{NN} W^{-1}_{NN}-1)\ )(2W+\zeta^*M_f)^{-1}_{Nm}
 \end{eqnarray}
 Where
 \begin{eqnarray}
 \Delta&=& 2-\zeta^*(1-W_{NN}.W_{NN}^{-1})\\
 \Delta '&=& 2\big[ ( 1+W_{NN}.W_{NN}^{-1})+(1-W_{NN}.W_{NN}^{-1})\cos(\frac{2\pi j}{n})\big]
 \end{eqnarray}

 $W_{NN}^{-1}\equiv(W^{-1})_{NN}$ is the matrix element of  $W^{-1}$ .

 Let us note that  we so far have not assumed anything about $N$ and all these results are valid for arbitrary $N$.

 Now it is a matter of a length but straightforward calculation to compute the trace
 \begin{eqnarray}
 \mathrm{Tr}(\mathbb{V}_-^{-1}-\mathbb{V}^{-1}_+)(\mathbb{W}_--\mathbb{W}_+)(\mathbb{W}_-+\mathbb{W}_+)^{-1}=8\sum_{j=1}^{n-1}\a_+\a_-\frac{(W_{NN}^{-1})^2-V_{NN}^{-3/2}W_{NN}}{\Delta'}
 \end{eqnarray}

 then it follows

 \be\label{renyi3}
 \ln\mathrm{Tr}\rho_A^n=- \sum_{j=1}^{n-1} \a_+\a_- \int_{0}^{\infty} ((W_{NN}^{-1})^2-V_{NN}^{-3/2}.W_{NN}) \frac{dE}{\Delta'}
 \ee

 We note here that it is already  possible at this stage to extract the divergent piece of the entanglement entropy  in the limit   $N\rightarrow \infty,~~ \mu \rightarrow 0$ by expanding    the integrand in the above equation around $E=0$ and using the asymptotic behavior of $ W_{NN}, W_{NN}^{-1} $ and $ V_{NN}^{-3/2}$ (see below). However, it turns out that we can do better than this and obtain an exact compact expression for EE of one-point in $1+1$.

 To that end it is convenient to introduce the following change of variable $\omega (E)= W_{NN}.W_{NN}^{-1}$.  Then eqt (\ref{renyi3}) becomes

 \begin{eqnarray}\label{entronepoint}
 \ln\mathrm{Tr}\rho_A^n &=&-\frac{1}{2} \sum_{j=1}^{n-1} (1-\cos(\frac{2\pi j}{n}) ) \int_{1}^{\omega_0}  \frac{d\omega}{(1+\omega)+(1-\omega)\cos\frac{2\pi j}{n} }\nonumber \\
 &=& \frac{1}{2}  \sum_{j=1}^{n-1} \ln ( (1+\omega_0)+(1-\omega_0)\cos\frac{2\pi j}{n}) =\ln [(C+1/2)^n-(C-1/2)^n ]
 \end{eqnarray}
 where $C =\frac{\sqrt{\omega_0}}{2}$ and $\omega_0= W_{NN}^{-1} W_{NN} | _{E=0}$.

 It is interesting to note to note that equation (\ref{entronepoint})  is exactly what one should have expected to obtain from the beginning.  The final result  we have obtained is only a rediscovery of the formula obtained in \cite{Peschel2003} for the case of one point entanglement in terms of two-point correlator functions. To bring this out more explicitly  we note that the relevant  correlation functions of the field variable  and its conjugate  are given in this case by
 \be
 X_{NN}=<\phi_N \phi_N>=\frac{1}{2} W_{NN}^{-1}|_{E=0}= \frac{1}{2} (V^{-1/2})_{NN}
 \ee

 \be
 P_{NN}=<\pi_N \pi_N>=\frac{1}{2} W_{NN}|_{E=0}= \frac{1}{2} (V^{1/2})_{NN}
 \ee

 It therefore follows that $ C= \sqrt{X_{NN} P_{NN}}$.

 Consider now the limit $N\rightarrow \infty$. 		

 \begin{eqnarray}
 W_{NN}|_{E=0} &=&\frac{1}{2\pi} \int_{0}^{2\pi} ( \mu^2+2-2\cos q)^{1/2} dq =  \frac{(a_0^2-1)^{1/4}}{2^{1/2}}P_{1/2}(\frac{a_0}{\sqrt{a_0^2-1}}) \nonumber \\
 W_{NN}^{-1}|_{E=0}&=& \frac{1}{2\pi} \int_{0}^{2\pi}(\mu^2+2-2\cos q)^{-1/2} dq= \frac{2^{1/2}}{(a_0^2-1)^{1/4}} P_{-1/2}(\frac{a_0}{\sqrt{a_0^2-1}}) \nonumber \\
 ~~~a_0 &=& \mu^2/2+1
 \end{eqnarray}

 $P_{1/2} $ and $P_{-1/2}$ are the associated Legendre functions of the first kind.

 In the continuum limit $\mu=m\epsilon \rightarrow 0$  and using the asymptotic form of $P_{1/2} $ and $P_{-1/2}$  we find
 $$
 C=\frac{2}{\pi} (-\ln\mu)^{1/2}
 $$

 and we finally obtain
 \be\label{dlog}
 \ln\mathrm{Tr}\rho_A^n= (n-1)(\frac{1}{2}\ln( -\ln m\epsilon) +\ln\frac{2}{\pi} )+\ln n
 \ee

 Therefore our calculation confirms the universal character of the divergence already observed in the finite interval case, however the origin of divergence in our case is slightly different although not fundamentally. Whereas in the finite interval case the divergence shows up in the massless limit, $m \rightarrow 0$, in  our case it arises as   the size of the region shrinks  to zero, becoming of the order of the UV cutoff. Nevertheless, the divergences in both cases has the same heuristic interpretation given in  \cite{Casini:2009sr},  it is due to the development of zero modes in the limit $\mu  \rightarrow 0$, when taking $\epsilon$ to zero, as can easily be seen from the potential matrix and eigenvalues, eqts (\ref{p}) and (\ref{potential1}), which in turns leads to a logarithmic divergence in the two-point correlation function, $W_{NN}^{-1/2}|_{E=0}$.  This coupling between the IR behavior or the zero mode  and the UV  cutoff in the massless limit of $1+1$  scalar theory  is a distinctive property that is worth discussing in this connection. 

 We first note that while the UV divergences of EE in $1+1$ scalar theory are widely discussed,
 IR diveregnces are much less studied. A  focused numerical discussion of the zero modes and IR diveregences  for $1+1$ scalar theory can be found in \cite{Yazdi:2016cxn}, see also \cite{Chandran:2015vuk} for more general and related  discussion  .

Second,  it is  well known that $1+1$ massless scalar theory has no normalizable ground state  in the full Minkowski space  nor in  $ R \times S^1 $ - like the one we considered in our paper by imposing periodic boundary conditions.  Therefore if we start with a massive scalar theory in Minkowski space or with its lattice version, impose periodic boundary condition ( $R\times S^1$), compute the EE and take the massless limit we should not be surprised to get IR divergences.  However  what seems to be more interesting is the consequences that this IR divergences has on the UV scaling of EE. 

 This coupling between the UV and IR cutoffs can be traced back to what may be referred to as the IR nonlocality that accompanies massless field theories in $1+1$ dim \cite{Chandran:2015vuk}. This IR nonlocality has two related symptoms. The Green function of a massless scalar field grows logarithmically rather than falling off with separation, and the absence of a normalized ground state in both full Minckowski space and $R \times S^1 $. 
 
 The concequenses of this coupling between the two scales of the theory can now be used to heuristically understand how the double-logarithmic divergence we obtained in (\ref{dlog}) emerges as a concequence of the developement of zero mode in the limit $\epsilon \rightarrow 0$ instead of $m \rightarrow 0$.
 
 If  we tarce out an interval of order of   UV cutoff,  one lattice spacing say, in this  case one would expect the usual universal logarithmic divergent term, $ \frac{1}{3}\ln \frac{L}{\epsilon}$ , to become of order one, whereas the double-logarithmic   becomes instead a leading term. Of course the this double-logarithmic term blows up not as a consequence of $m\rightarrow 0$, which we kept finite,  but as we mentioned above  it is rather due to the fact that the size of the interval has become of the order of $\epsilon$. 
 
To see  how this is related to the IR divergence of the theory  we note that for an  infintesimal interval  and based on dimensional grounds we are left with    two scales at our disposal in the theory, the mass $m$ or the IR cutoff, and $\epsilon$ or $L$ (but not both $L$ and $\epsilon$, the length of the interval is no longer an independent scale as it is of order of $\epsilon$). Therefore  the entropy will be controlled by  the dimensionless parameter $ m\epsilon=\mu \sim m L$, and it is this coupling between $m$ and $L$ or $\epsilon$ which brings in the same divergence we see in the massless limit. Technically speaking there is no way to distinguish between taking $m$ to zero or $\epsilon$ to zero. The situation is somehow similar to the case of an infinite half line in 1+1,  in which the theory  only has two scales and they are coupled to give one dimensionless parameter $\mu=m\epsilon$. The potential, or more precisely the rescaled potential,  depends only on this dimensionless parameter and it develops zero modes whether we take $m$ to zero or $\epsilon$ to zero. The EE is a function of $m\epsilon$, and the logarithmic scaling of the entropy, $\ln m\epsilon$, can intuitively be related to the IR behavior of the massless theory and give a heuretsic explanation to why the area law in $1+1$ is logarithmic  rather than being just a constant as one would have naively expected, see for instance \cite{Chandran:2015vuk} .

 Now, if  we instead considered an interval made of finite number of points, the size of the interval would be still infinitesimal and we would obtain an EE with the following behavior

 \be
 \ln\mathrm{Tr}\rho_A^n\sim (n-1)\frac{1}{2}\ln( -\ln m L)  +\cdots
 \ee
 where $L=p \epsilon$, $p$ the number of points making the interval.

 We shall not try to expand the discussion about the physical relevance of considering the entanglement of an infinitesimal interval, which may loosely speaking be seen as simulating a black hole in its final faith, in one 1+1, because no one would expect  our picture of space-time and field theory will continue to make sense when the size of the black hole becomes of the order of the Planck scale say. The relevance of our result is limited to showing the universal character of a softer IR/UV  divergence  using  different regularization scheme and different technique  .

 Before concluding this section,  there is the question of how the current approach can be used to compute EE for a finite interval in 1+1. We shall not try to tackle this problem, however  we believe that the case of one point already sheds some light on the technical challenge that one would face. The main mathematical difficulties is finding the inverse of $ (\mathbb{W}_-+\mathbb{W}_+)^{-1}$. Although partial (block) inversion seems to be always possible, the remaining $N\times N$ matrices are generally not easy to invert. In the one-point case it turned out that one has  to invert a matrix given by the sum of a circulant matrix and two rank-one matrices, and therefore it was easy to invert the full matrix using Miller theorem. Now, it is not difficult to see that if we considered an interval made of $p$ points say, the resulting matrix would be a sum of circulant matrix and two (may be more) rank-$p$ matrices and here too Miller theorem can be efficiently used, as $p$ will be much smaller than $N$ ( $p \ll N$, or $L\ll R$ ). The addition of new points, $p$ points say, will increase the entropy;  for the interval to be of a finite size $p$ should be of the order of $\frac{1}{\epsilon}$ and this is how one would expect the universal  divergence $\sim \ln p=  \ln \frac{L}{\epsilon}  $ to build up.

 \section{Interacting Theories, Beyond The Gaussian State}
 As emphasized in the introduction one of the main motivation for developing  Green's function approach for R\'{e}nyi entropy for models with finite number of degrees of freedom (  lattice or fuzzy models) is to investigate the effects of interaction  on the behavior of EE. Indeed free field theories and Gaussian models are generally an approximate idealization of  more realistic models for nature. In this section we shall show how  the present technique is used to systematically compute the correction due to interaction (quartic type) order by order in perturbation theory.

 Let us consider $\lambda \phi^4$ type interaction added to  our starting lagrangian of equation (\ref{LagStand}).  We first note  that  this quartic interaction can be introduced in two different natural  ways. The first is a lattice    discretization of  $\phi^4$ theory,  the lagrangian (\ref{scalarlag}), which leads to  an interaction term of the form  $\sum q_i^4$. The second possible form is ${\lambda} (Q^TQ)^2$,  which can not be interpreted  as resulting from the discretization of a local interaction term as  we  shall show below. The  two different interaction terms will be shown to lead to drastically different results.

\subsection{Local Interaction}	
 Let us start by the following lagrangian

 \be\label{lagint}
 L_E=\frac{1}{2}( \dot{Q}^T \dot{Q} +Q^T V Q)+\frac{\lambda}{4!} \sum_{i=1}^{N} q_i^4
 \ee

 This is just a lattice discretization of the $1+1$ interacting scalar theory .

 In view of the fact that the interaction term is invariant under the action of the permutation matrix, a perturbation series for $\ln \mathbb{Z}_n(\lambda)$ is straightforward to write down and formally similar  to the calculation of \cite{Hertzberg:2012mn} in the continuum.

 \be\label{enper}
 \ln \mathrm{Tr}\rho_A^n(\lambda)= \ln \mathbb{Z}_n(\lambda)-n\ln Z_1(\lambda)
 \ee

 A generating partition function $\mathbb{Z}_n(J) $ can easily  be introduced to establish  Wick's theorem for this formalism and obtain  the following leading order correction to R\'{e}nyi entropy \footnote{The Green's function  is  understood here to be evaluated at $E=0$.}

 \be\label{enper1}
 \ln \mathrm{Tr}\rho_A^n(\lambda)= \ln \mathbb{Z}_n(0)-n\ln Z_1(0)-\frac{3\l}{4!}\int d\t \big[  \sum_{i=1}^{nN}\mathbb{G}_{ii}(\t,\t)^2 -n \sum_{i=1}^{N} G_{ii}(\t,\t)^2\big ]+ \mathcal{O}(\l^2)
 \ee

 As an explicit example let us reconsider the case of 2-coupled H.O and compute the first order correction to the entanglement entropy due to an interaction quartic term of the local type.

 Considering the  correction due to quartic interaction  for 2-couple H.O should furnish  a good pedagogical exercise for developing a perturbation theory  that allows one to compute the effect of interaction order by order; however in view of the  limitation of the real time approach to Gaussian models this problem has to our knowledge never been tackled systematically. On the other hand we shall  use this   interacting model ( non-gaussian) to address  the issue of the space-time nature of the EE. This issue was raised by Sorkin in the context of  continuum field theory in order to free the entanglement entropy from reference to a density matrix localized to hypersurface \cite{Sorkin:2012sn} . A covariant  definition of spacetime entropy ( which includes entanglement entropy)  was derived using non other than the Peirls bracket and the related spacetime two-point correlation functions. This covariant definition was  limited to gaussian states or free field theories.  Related to this definition there is the well known result  for bosinic and fermionic lattice systems, where it was shown that reduced  density operator can be determined from the properties of the two-point correlation functions for gaussian models \cite{Peschel2003}.

 Motivated by the  work of Sorkin the authors of \cite{Chen:2020ild} tried to generalize the results of \cite{Sorkin:2012sn} and investigated the existence of a covariant spacetime definition  of entropy for non-gaussian  states or interacting theories. Therefore we shall start by  briefly reviewing their main results and put them in the context of the model that we will consider below.

 Consider first a generic gaussian state $\rho(q,q')$ for one  degree of freedom\footnote{The field theory problem can be divided into a series of calculations  each involoving a single degree of freedom \cite{Sorkin:2012sn}.},

 \be\label{densityguaussian}
 \sqrt{\frac{A}{\pi}} e^{-A/2(q^2+q'^2) -C/2 (q-q') ^2 }
 \ee

 The the different correlators can be computed using this density.

 $$
 <\hat{q}\hat{q}> =1/2A, ~~~ < \hat{p}\hat{p}>= A/2 +C,~~~<\hat{q}\hat{p}>=<\hat{p}\hat{q}>^* = i/2
 $$
 It was convinient to define
 $$
 \sigma^2 =<\hat{q}\hat{q}><\hat{p}\hat{p}>-(Re<\hat{p}\hat{q}>)^2=1/4 +C/2A
 $$
 and
 $$
 \xi = \frac{2\sigma -1}{2\sigma+1}
 $$

 A standard results  can be used to compute the associated entropy
 \be\label{entropysorkstud}
 S= -\mathrm{Tr}\rho \ln \rho = -\ln (1-\xi) -\frac{\xi}{1-\xi}\ln\xi
 \ee

 This result can as well   be  obtained using the Replica Trick by computing  first $\mathrm{Tr} \rho^n$.

 \be
 \mathrm{Tr} \rho^n =\frac{(1-\xi)^n}{1-\xi^n}
 \ee	

 using $S= \lim_{n\rightarrow 1} \partial_n \mathrm{Tr} \rho^n$  equation (\ref{entropysorkstud}) is recovered.

 In \cite{Chen:2020ild}  a conjecture regarding the behavior of the entanglement entropy to the first order in perturbation theory was put forward and justified by explicit calculation for a particular non-gaussian state.

 Consider for instance the following non-gaussian state

 \be\label{nong}
 \rho_\l (q,q')= Ne^{-\frac{A}{2}(q^2+q'^2)-\frac{C}{2}(q-q')^2 -\l [\a_1(q^4+q'^4)+\a_2(q^3q'+qq'^3)+\a_3 q^2 q'^3]}
 \ee

 $N$ is a normalization constant, $\l$ is the interaction constant, the perturbation parameter.

 We first mention that we shall show exlicitly  using the Green's function approach that this non-gaussian state does actually simulate the reduced  density operator for 2-coupled H.O with quartic interaction.

 Let us conjecture that to the first order in $\l$ the entropy formula in the Gaussian case, eqt(\ref{entropysorkstud}), keeps the same form but with the correlators are now evaluated with respect to non-guassian state $\rho_\l$.

 Those correlators were computed in \cite{Chen:2020ild} and the modified $\xi_\l$ was shown to be given by

 \bea\label{xil}
 \xi_{corr} &=& \frac{2\sigma_\l -1}{2\sigma_\l+1} \nonumber \\
 &=& \xi + \l \bigg[ \frac{3\xi }{(\xi+1)(\xi-1)^3}\a_1 +\frac{3(\xi+1)}{2(\xi-1)^3\beta^2} \a_2 +\frac{1+\xi +\xi^2}{(\xi+1)(\xi-1)^3 \beta^2 }\a_3\bigg]
 \eea

 where $\beta= \frac{1}{2} ( A \sqrt{1+2C/A}+A+C)$.

 On  the other hand we can compute the entanglement entropy directly by the replica trick using $\rho_\l$ .

 This  gave to  the first order in $\l$ the following entropy

 \bea\label{entreplica}
 S &=& -\ln (1-\xi) -\frac{\xi}{1-\xi}\ln\xi -\l \frac{3\xi\ln \xi}{(\xi+1)(\xi-1)^5 \beta^2} \a_1 \nonumber \\
 &-& \l \frac{3(\xi+1)\ln\xi}{2(\xi-1)^2 \beta^2} \a_2-\l \frac{3(\xi^2+\xi+1)\ln\xi}{(\xi+1)(\xi-1)^5 \beta^2} \a_3
 \eea
 It  is now easy to show that the above expression of the entropy is none other than the expansion of the following formula to the first order in $\l$
 $$
 S= \ln (1-\xi_{corr}) -\frac{\xi_{corr}}{1-\xi_{corr}}\ln\xi_{corr}
 $$

 Therefore we reach the conclusion that EE standard formula (\ref{entropysorkstud}) for Gaussian states holds to the first order  in perturbation theory beyond the Gaussian theory and EE is still captured by  the two-point  correlators (the perturbed ones).

 Of course the above procedure does not  for instance allow one to compute systematically the correction to EE  for interacting theory. It only shows that to the first order in perturbation theory the EE is given by the same formula valide for the Gaussian state but with the correlators replaced by their non-gaussian counterparts. Actually this  was shown in  \cite{Chen:2020ild} to be a  general future of any perturbation to the first order.

 In what follows we shall  work out an explicite example, generic enough, of interacting theory with quartic interaction  and compute from the first principle the first order correction to  EE and compare our results with that of \cite{Chen:2020ild}.

 The lagrangian is given by

 \be\label{intHO}
 L= \frac{1}{2} [ \dot{q}_1^2 +\dot{q}_2^2+ \omega^2 q_1^2 +\omega^2 q_2^2 +2 bq_1 q_2 ]+\frac{\l}{4!}( q_1^4 +q_2^4)
 \ee

 The gaussian part of this Lagrangian is basically the model we have previously seen, equation (\ref{potentialHO}), with matrix potential given by

 \be
 V= 	
 b\left(
 \begin{array}{cc}
 	a &1 \\
 	1 &a \\
 \end{array}
 \right),~~~~~ a=\omega^2/b,
 \ee

 Its free, $\l=0$, reduced  density matrix $\rho(q,q', \l=0)$ is of the from given by equation (\ref{densityguaussian}). For $\l \neq 0$ the density matrix is unknown,  however within our approach based on Green's function we do not need to construct the  reduced density matrix explicitly to compute the correction to EE.

 According to (\ref{enper1}) the correction to  R\'{e}nyi  entropy is given  to the first order in perturbation theory  by
 \be
 \delta \ln \mathrm{Tr}\rho^n(\lambda ) =-\frac{3\l}{4!}\int d\t \big[  \sum_{i=1}^{2n}\mathbb{G}_{ii}(\t,\t)^2 -n \sum_{i=1}^{N} G_{ii}(\t,\t)^2\big ]+ \mathcal{O}(\l^2)
 \ee

 In appendix II we show that

 \bea\label{corr1}
 \int d\t \big[  \sum_{i=1}^{2n}\mathbb{G}_{ii}(\t,\t)^2 -n \sum_{i=1}^{2} G_{ii}(\t,\t)^2\big ]&=& \frac{1}{b\sqrt{b}}\bigg[ n[ -\frac{3}{16\l_+ \l_-} -\frac{3}{16\l_+^2} -\frac{1}{4\l_+^2\l_-} -\frac{1}{4\l_+\l_-^2} \nonumber \\
 +\frac{1}{4\l_-\l_+(\l_++\l_-)}]    &+& [\frac{1}{2\l_+\l_-} + \frac{1}{4\l_+^2}-\frac{1}{2\l_-(\l_++\l_-)} ]B_-\nonumber \\
 +[\frac{1}{2\l_+\l_-} + \frac{1}{4\l_-^2}-\frac{1}{2\l_+(\l_++\l_-)} ]B_+ &+& \frac{1}{n} [\frac{1}{4\l_+} B_-^2+ \frac{1}{4\l_-} B_+^2+\frac{1}{\l_-+\l_+} B_- B_+ ]\bigg]
 \eea

 where
 $$
 \l_{\pm}= \sqrt{a\pm1}, ~~B_\pm=\frac{1}{2}\sum_{j=1}^{n-1} \frac{2\r\pm \sigma(1+\cos\frac{2\pi j}{n})}{2\r^2-\s^2(1+\cos\frac{2\pi j}{n})},~~ \r=\frac{\l_++\l_-}{2}, ~~~\s= \frac{\l_+-\l_-}{2}
 $$

 $B_\pm$ can be summed using standard summation formula. $\l_\pm$ can be expressed using $\xi$ as defined in the previous section, equations (\ref{reyni})  and (\ref{reyni2}).

 Using the definition of $\xi$  it is a straightforward exerecise to express the leading order correction to the Reyni entropy   $	\delta \ln \mathrm{Tr}\rho_A^n(\lambda )$ as

 \be\label{enper2}
 \delta \ln \mathrm{Tr}\rho_A^n(\lambda)=-n\frac{\l}{32 b\sqrt{b}}(x^2-1)^{3/4}\sqrt{2x}[ x^2(1-12x^2)+2x(4x^2-1)B_n+ (4x^2+1) B_n^2]
 \ee

 where  $B_n=\frac{1+\xi^{n}}{1-\xi^{n}}$, $x=B_1= \frac{1+\xi}{1-\xi}$.

 As can be easily checked the correction vanishes for $n=1$ and for $\xi =0$ or $x=1$. The vanishing of $\delta \ln \mathrm{Tr}\rho_A^n(\lambda )$ for $\xi =0$ reflects the fact that 	quartic interaction chosen induces no new entanglement and the correction is due to the perturbation of the ground state.

 The correction to the entanglement entropy is obtained by the replica trick
 \be
 -\frac{d}{dn}\delta \ln \mathrm{Tr}\rho_A^n|_{n=1}=\delta S_{ent}=  \frac{2\l}{b\sqrt{b}} \sqrt{1+\xi} (\frac{1+\xi}{1-\xi})^3 \xi^{7/4}  \frac{\ln \xi}{(1-\xi)^2}
 \ee

 This leading order correction to the entanglement entropy is very particular and confirms the results we discussed above about   the covariant spacetime definition of EE and the validity of Gaussian EE formula beyond the Gaussian states or for interacting  theories to the first order \cite{Chen:2020ild}.

 This can be explicitly brought out by noting that  $\delta S_{ent}$ can be obtained from the first order expansion of the following entropy expression

 \be\label{entnongauss}
 S_{ent}(\l)= -\ln (1-\xi_\l) -\frac{\xi_\l}{1-\xi_\l}\ln \xi_\l
 \ee
 where

 \be\label{xil2}
 \xi_\l= \xi +\delta \xi,~~~~ \delta \xi = \frac{2\l}{b\sqrt{b}} \sqrt{1+\xi} (\frac{1+\xi}{1-\xi})^3 \xi^{7/4}
 \ee	

Or for $\ln \mathrm{Tr}\rho_A^n$ we have

\be
\ln \mathrm{Tr}\rho_A^n= \ln \frac{(1-\xi_\l)^n}{1-\xi_\l^n}
\ee

 Therefore to the first order in perturbation theory EE is given by the same expression as EE at the free level with  the parameter $\xi$ being replaced by its perturbed counter part evaluated using the pertubed two-point correlators.

 In addition we can use our result to deduce the corresponding non-guassian reduced matrix $\rho_\l(q,q')$ to the first order by equating $\xi_\l$ of equation (\ref{xil2}) with $\xi_{corr}$ of equation (\ref{xil}) and hence fixing $\alpha_1, \alpha_2, \alpha_3$ appearing as arbitrary parameters in (\ref{nong}).

 The natural question that arises now  is what about higher orders?  Do the two points correlators continue to capture EE and  does the Gaussian standard formula for the entropy continue to be  valid beyond the first order of perturbation theory?

 According to a general argument presented in \cite{Chen:2020ild} higher order correlators start to contribute from the second order on. Therefore the validity of the Guassian formula for interacting theory seems to be a particular future of first order perturbation theory, possibly due to the fact that the tunneling terms of the Green's function, $\mathbb{G}_{\pm\mp}$, do not contribute at this order. Nevertheless it would be interesting to  push the calculation beyond the first order using our approach.    We hope to return to this point in the future.

 \subsection{Non Local Interaction}
 In  this subsection we consider  another type of quartic interaction which looks natural to be added to the free lagrangian of a set of coupled H.O. This interaction term is  of the form   $\frac{\lambda}{4} (Q^TQ)^2$.

 Using Wick's theorem it is straightforward to show that the first order correction in $\l $ is given by
 \be\label{enper3}
 \delta \ln \mathrm{Tr}\rho_A^n(\lambda)= -\frac{\l}{4}\int d\t \bigg[2\big ( \mathrm{Tr} \mathbb{G}(\t,\t)^2 -n \mathrm{Tr} G(\t,\t)^2) +(\mathrm{Tr} \mathbb{G}(\t,\t))^2 -n (\mathrm{Tr} G(\t,\t))^2\bigg ]
 \ee

 In this correction we can notice that there are two terms, the first comes from what we may refer to as connected Feynman diagrams and can be shown to be finite and formally equal to

 \be\label{foc}
 (\delta \ln \mathrm{Tr}\rho_A^n)_{c}= -\frac{\lambda}{2} \bigg[ -\frac{3n}{8} \mathrm{Tr}V^{-3/2}+\frac{1}{4} \mathrm{Tr}(\mathbb{V}_+^{-1}+\mathbb{V}_-^{-1})(\mathbb{W}_-+\mathbb{W}_+)^{-1} +2F(\mathbb{W}_-,\mathbb{W}_+) \bigg]
 \ee
 where $F $ is defined by
 $$
 F(\mathbb{W}_-,\mathbb{W}_+)= \int_{0}^{\infty} \mathrm{Tr}\big[ (\mathbb{W}_-+\mathbb{W}_+)^{-1} e^{-2\mathbb{W}_+\t} (\mathbb{W}_-+\mathbb{W}_+)^{-1}e^{-2\mathbb{W}_+\t}\big]
 $$

 The second contribution\footnote{This term was mistakenly omitted in \cite{Allouche:2018err} }$(\mathrm{Tr} \mathbb{G}(\t,\t))^2 -n (\mathrm{Tr} G(\t,\t))^2)$, which can be interpreted as coming from disconnected Feynman diagrams, has unremovable IR divergence coming from the integration over $\t$. More explicitly we have

 \bea\label{disc}
 \int_{-\infty}^{\infty} d\t [\mathbb{G}(\t,\t))^2 -n (\mathrm{Tr} G(\t,\t))^2) ] &=&  \int_{0}^{\infty} d\t \bigg[ \frac{1}{2}(\mathrm{Tr} \mathbb{W}_+^{-1}e^{-2\mathbb{W}_+\t }  )^2+ \mathrm{Tr}} \mathbb{W}_+^{-1}\mathrm{Tr}\mathbb{W}_+^{-1}e^{-2\mathbb{W}_+\t \nonumber \\
&+& \frac{1}{2}(\mathrm{Tr} \mathbb{W}_+^{-1} )^2- n \frac{1}{2}(\mathrm{Tr} {W}^{-1} )^2 \bigg]
 \eea

 The first and second term on the right hand side of the above equation give finite contribution. However  the last two terms do not cancel each other and lead to an IR unremovable infinity. Indeed unlike what happened  in the previous interaction type, $\sum q_i^4$,  or in the Gaussian case , where this kind  of IR divergence is cancelled out by the normalization term ($-n \ln Z$), here the two terms  scale differently  with $n$, namely we have,

 \be
 \int_{0}^{\infty} \bigg[\frac{1}{2}(\mathrm{Tr} \mathbb{W}_+^{-1} )^2- n \frac{1}{2}(\mathrm{Tr} {W}^{-1} )^2\bigg] = \frac{( n^2-n)}{2}(\mathrm{Tr} {W}^{-1} )^2  \int_{0}^{\infty} d\t =\infty
 \ee

 This IR divergence may seem surprising or unexpected. Actually it can be shown that the contribution of this kind of disconnected diagrams can be summed to all order . The natural question is what is peculiar about this interaction term?. To try to answer this question we go back to the interaction term $ (Q^TQ)^2$. If we expand this term we easily notice that it is a highly non-local,  each oscillator couples to all other oscillators in the chain,  and hence  can not be understood as arising from the discretization of a local interaction term of scalar theory, in other words it is a long-range  interaction and  induces  couplings between different socillators at the interaction level. Moreover this interaction or this model can not emerge from the discretization of a Lorentz invariant contiunuum theory.   
 
 It is important to note here that at the Guaussian level, quadratic lagrangian,  long-range  interaction or non-local terms in continuum field theories do not seem to  have this sort of IR problem. Long-range peotenials or non-local theories only change the scaling of the EE as one would expect, see \cite{Shiba:2013jja}  for an example of non-local field theory and  references to earlier works. The absence of this kind of IR divergence  at the Guassian level can in fact be seen from our main formula for the EE  given by equation (\ref{ent3}), which is valide for any positive definit potential, whether is a long-range or short -range.  However, when it comes to interactions
 our  result strongly suggests that EE does not tolerate non-local interactions, at least not any type of non-local interaction, or interaction which induces  higher order coupling between different oscillators.  In other words and technically speaking,   it is the fact that the chosen interaction  induces new entanglement, terms like $q_i^2 q_j^2$,  which  is the source of the problem upon the following Wick contractions :

 \begin{equation}
 \sum_{i,j}\wick{\c q_i \c q_i} \wick{\c q_j \c q_j}= (\mathrm{Tr} G)^2 
 	\end{equation}

 This type of contraction leads to an IR divergence due to the different scalings  of the $n$-fold Green's  matrix function and $n=1$ Green function after taking the trace, unlike local interaction where the oscillators degrees of freedom are not coupled by the interaction and no such contraction or disconnected diagrams would be present. It is important here to note that by local interaction we refer only to the type considered in the previous section, i.e $\sum q_i^4$, and this leads us to the following natural question. Is this IR divergence associated with the non-locality of the quartic interaction, i.e. long-range, or is it a symptom that will accompany any interaction which couples different oscillators, even if it is a short-range (nearest neighbor quartic interaction)\footnote{Short-range quartic interaction could be seen as emerging from a lattice discretization of a term like $(\partial_x \phi)^4$.}  
 
 Although we do not have a  definite general answer for this question, we believe that it is the case for all one dimensional chains with quartic interactions which couple different oscillators. However, for other non-local models such as  $2+1$ fuzzy field theories the story seems a bit different. An outlook on the  EE for interacting fuzzy scalar theories is the subject of the next and last section of this paper

 \section{Applications to Fuzzy Spaces and Outlook}

The main motivation behind the development of this discrete Green's function approach to EE is to investigate EE on fuzzy spaces, in particular the effect of interaction in connection with IR-UV mixing phenomena on EE. Indeed in view of the IR divergence we met when we considered a particular quartic interaction and the phenomena of UV-IR mixing, it is not a priori obvious how the entanglement entropy will respond to a natural quartic interaction on fuzzy or non-commutative underlying geometry. Although this issue is the project of future work  and shall not be  tackled  in the present work, we are going to briefly outline how field theories on fuzzy spaces  fit into this framework and the possible challenges to analytically investigate them . To that end we consider scalar theory on $1+2$ space-time, with space modeled by a fuzzy sphere.

 The entanglement entropy on fuzzy sphere was first considered in \cite{Dou:2006ni,Dou:2009cw} and later  by several authors \cite{Karczmarek:2013jca,Sabella-Garnier:2014fda} .

 The starting Lagrangian in \cite{Dou:2006ni} was

 \be\label{fuzzylag}
 L=\frac{1}{2}Tr \big(\dot{\phi}^2
 +\phi\big[{\cal L}_i^2-{\mu}^2\big]\phi \big)
 \ee

 The scalar field $\phi$ is an $N{\times}N$ hermitian matrix with
 mass parameter $\mu$. The Laplacian ${\cal L}_i^2$ is the $ SU(2)$
 Casimir operator given by ${\cal L}_i^2={\cal L}_1^2+{\cal
 	L}_2^2+{\cal L}_3^2$ with action defined by ${\cal
 	L}_i({\phi})=[L_i,\phi]$ and ${\cal
 	L}_i^2({\phi})=[L_i,[L_i,\phi]]$. The $L_i$ generate the
 $SU(2)$ irreducible representation of spin $l=\frac{N-1}{2}$.

 By using the canonical basis where $L_3$ is diagonal and choosing suitable field variables it was possible to show that
 the above free scalar theory on a fuzzy sphere  into  splits naturally into $2(2l)+1$ independent sectors $
 \{\mathcal{H}_m\} , m=-(N-1),\cdot\cdot\cdot\cdot, (N-1)$, each
 sector $\mathcal{H}_m$ has $N-|m|$ degrees of freedom ($N-|m|$
 coupled harmonic oscillator) and described by a Lagrangian $L^{(m)}$. More explicitly we have
 \be
 L=\sum_{-(N-1)}^{N-1}L^{(m)}=
 \sum_{m=-(N-1)}^{N-1}\frac{1}{2}[\dot{Q}_m^T \dot{Q}_m-Q_m^TV^{(m)} Q_m]
 \ee

 where
 \begin{eqnarray}
 V_{ab}^{(m)}=2\bigg[\big(c_2+\frac{{\mu}^2}{2}-A_aA_{a+|m|}\big){\delta}_{a,b}
 -\frac{1}{2}B_{a-1}B_{a-1+|m|}{\delta}_{a-1,b}-\frac{1}{2}B_{a}B_{a+|m|}{\delta}
 _{a+1,b}\bigg].\label{V}
 \end{eqnarray}

 and  $ B_a = \sqrt{a(N-a)}$ and $ A_a =-a+\frac{N+1}{2}$,  $a,b= 1\dots  N-|m|$.

 The new canonical variables $Q_a^{(m)}$ are introduced after decomposing  $\phi$ as $ \phi=\phi_1+i\phi_2$ and defining $ \Phi=  \phi_1+\phi_2$.

 \begin{eqnarray}\label{cv}
 Q_m=(
 \Phi^{1,1+m}, \Phi^{2,2+m}, \cdot  \cdot \cdot ,\Phi^{N-m,N}
 )~,~Q^{(-m)}=(
 \Phi^{1+m,1}, \Phi^{2+m,2}, \cdot  \cdot \cdot ,\Phi^{N,N-m}
 )
 \end{eqnarray}

 In order to introduce the entanglement entropy for these fuzzy models one needs first to divide the field degrees of
 freedom into two sets residing in two, or several, disjoint regions.
 In \cite{Dou:2006ni} it was conjectured using heuristic arguments that the field variables represented  by the elements of the upper left triangular part of the field matrix
 $\Phi_+$  and the right lower triangular part    $\Phi_-$  correspond   respectively  to the upper and lower fuzzy hemispheres. This conjecture was subsequently  proven in \cite{Karczmarek:2013jca} using  rigourous mathematical argument.

 More explicitly for $N=5$ we have

 \begin{eqnarray}
 \Phi_{+}=\left (\begin{array}{ccccc}
 {\Phi}_{11} & {\Phi}_{12} & {\Phi}_{13} & {\Phi}_{14} & 0 \\
 {\Phi}_{21} & {\Phi}_{22} & {\Phi}_{23} & 0 & 0 \\
 {\Phi}_{31} & {\Phi}_{32} & 0 & 0 & 0 \\
 {\Phi}_{41} & 0 & 0 & 0 & 0 \\
 0 & 0 & 0 & 0 & 0
 \end{array}\right)~,~
 \Phi_{-}=\left (\begin{array}{ccccc}
 0& 0 & 0 & 0 & \Phi_{15}\\
 0 &0 & 0 & \Phi_{24}  &  \Phi_{25}\\
 0 & 0 & \phi_{33}  & \phi_{34} & \Phi_{35}  \\
 0 &  \Phi_{42} & \Phi_{43} & \Phi_{44}  & \Phi_{45} \\
 \Phi_{51}& \Phi_{52} & \Phi_{53} & \Phi_{54} &  \Phi_{55}
 \end{array}\right)
 \end{eqnarray}

 In terms of the canonical field variables $Q_m$ tracing out the upper or the lower hemi-fuzzy sphere corresponds to taking each individual sector $ \mathcal{H}_m$ and tracing out half of the degrees of freedom in each sector.

 In view of the fact that the different sectors do not mix for the free theory, the computation of the  R\'{e}nyi  entropy reduces to computing the EE in each sector selectly. Each sector has a Lagrangian of the standard form (\ref{LagStand}). The global ground state operator is given by
 \begin{eqnarray}
 \rho=\bigotimes_{m=-(N-1)}^{N-1} \rho^{(m)}.
 \end{eqnarray}

 and the reduced  one $ \rho_{+}=Tr_{-}\rho$ is given by
 \begin{eqnarray}
 \rho_{+}=\bigotimes_{m=-(N-1)}^{N-1} \rho_+^{(m)}.
 \end{eqnarray}

 where $  \rho_+^{(m)}$ is obtained by tracing out the degrees of freedom corresponding to the ones in the lower hemi-fuzzy sphere in each sector.

 In \cite{Dou:2006ni} it was shown numerically and using  the Real Time approach that the resulting EE is proportional to the number of freedom residing at the boundary. The result could  as well be interpreted as proportional to the circumference of the separating region, a fuzzy circle.

 The present Green's function technique can  similarly be applied to compute R\'{e}nyi  entropy (at least numerically). Indeed, in view of the canonical splitting of the Hilbert space of the free  theory, the  Green's function matrix for this model is given by

 \be\label{greenff}
 \mathbb{ G}_n(\t,\t')= \bigoplus_{m=-(N-1)}^{N-1} \mathbb{ G}^{(m)}_n(\t,\t')
 \ee

 Unlike the case of $1+1$ lattice the potentials $V^{(m)}$  are not circulant matrices, which poses  an additional complication to the analytical investigation of this model, and for the fuzzy disc as well \cite{Dou:2006ni}. However, we note that it was shown in \cite{Dou:2009cw} that  the EE is essentially captured by the near boundary approximation   where the potentials are   Toeplitz matrices. In addition to this it can be shown that not all sectors,  different $m$, are equally relevant, there are sectors which are weakly entangled and hence give  negligible contributions to the entanglement entropy in the large $N$ limit \cite{FatenThesis}, and the relevant sectors are governed by Toeplitz potentials in the near boundary approximation . Putting all this together it is conceivable that in the large $N$ limit one can approximate the potentials for the relevant sectors with circulant matrices which would greatly simplify the problem.

 Now, as we mentioned  this technique is specially devised  to go beyond the free gaussian theories 																																										 and investigate the effects of interaction on the EE on fuzzy spaces. Therefore we conclude this paper by outlining how this could be done.

 Let us assume that we have added a quartic interaction term to our starting lagarangian (\ref{fuzzylag}),
 \be\label{fuzzyint}
 V_{int} =\frac{\l}{4} Tr \phi^4
 \ee

 In order to apply the Green's function formalism the interaction term should expressed using the $\Phi$ or the new canonical variables $Q^{(m)}$.  We shall not go into the detail of this, however we shall mention few important points.

 First it is not difficult to see that this interaction term mixes different sectors and introduces new couplings between the degrees of freedom which were not coupled at the free level. Second  and related to the first the interaction term once expressed using the $n$-fold variables is not invariant under the action of the permutation matrix $ P_\pi$ within each sector.  Therefore the application of Wick's theorem is a bit subtle.

 Some natural questions arise here and  make the study of the above interaction worth considering. The first is how such interaction would enhance the gaussian entanglement in view of the UV-IR mixing phenomean related to both the interaction and the entanglement entropy in the large $N$ limit. Actually divergences of UV-IR mixing origin had already been observed in \cite{Dou:2006ni} at the free level in 2+1   . A second question concerns a possible unremovable IR infinity, for finite $N$,  similar to the one we met in the previous section  for the non-local interaction.

 As for the second question, i.e. IR divergence, we do not expect such unremovable IR divergence to be present in this model. Although the interaction term once expressed using the above canonical variables, eqt (\ref{cv}), seems to behave like the non-local term we considered in the previous section, this behavior is partially apparent. For instance the mixing of different sectors which may cause this infinity is similar to the mixing of different Fourier modes for any local interaction in standard continuum field theory, such a mixing do not show up at the free level. Nevertheless the interaction term (\ref{fuzzyint}) can be seen to introduce quartic coupling between different oscillators DF within the same sector ( the same $m$) and therefore only explicite calculation can completely settle this issue. Actually working the case $N=2$ would be enough, because as long as this particular IR problem is concerned the size of the fuzzy sphere should play no role.

  Now, in order to carry out an analytical study of the EE for free and interacting fuzzy field theories,     we believe that some mathematical tools should first be developed,  in particular understanding the asymptotic behavior of Toeplitz matrices with $N$-dependent index on which very little if any is known.

 Finally, we note that it is conceivable that one may have to take a different path, based on a different decomposition of the field variable, in order to analytically tackle the fuzzy regularization  using the Green's function technique. For instance we may reformulate the Green's function developed in the first section   by re-expressing the Euclidean path integral formula for the reduced density matrix using a doublet field  notation as follows.

 We let $\Phi=  \left( \begin{array}{c}
 \Phi_+ \\
 \Phi_-
 \end{array}
 \right)$. With this doublet notation the lagrangian (\ref{fuzzylag}) takes the form

 \be
 L=\frac{1}{2}Tr \big(\dot{\Phi}^T \dot{\Phi}
 -\Phi^T V \Phi \big), ~~~ V=\left (\begin{array}{cc}
 	{\cal L}_{++}^2-\mu^2 & {\cal L}_{+-}^2 \\
 	{\cal L}_{-+}^2 & {\cal L}_{--}^2 -\mu^2
 \end{array}\right)
 \ee

 where ${\cal L}_{\pm\pm}$ and $ {\cal L}_{\pm \mp}$ follow from the adjoint actions of the angular momentum operators, viz,
 $$
 \Phi_{\pm}{\cal L}^2(\Phi_{\pm}) =\Phi_{\pm}{\cal L}_{\pm \pm}^2\Phi_{\pm},~~...\text{etc}
 $$

 This formally will reduce the problem to a $2$-dimensional one with reduced Laplacians and one may hope to obtain a Green's function matrix that makes the problem more tractable analytically. As a by product of the formulation it is possible that one would end up deriving the Laplacian of $n$-fold cover of a fuzzy space and understanding the constructions of new class of fuzzy spaces .
 We hope to come to this and other related issues in future work.


 \acknowledgments

We are particularly  thankful to A.Botthcher for several very useful discussions on the asymptotic behavior of Toepltiz matrices with $N$-dependent symbol.

 We thank P. Calabrese for useful discussions.

We thank R.Sorkin for useful discussions and in particular for bringing to our attention  reference \cite{Chen:2020ild}.

 D.Dou would like to thank the High Energy section of ICTP for their kind hospitality during different stages of this work.

 This work is supported in part by the Algerian Minsitry of Higher Education and Scientific Research, Project R.F.U B00L02UN390120190003.

\section{Appendix I}

\subsection{Series Expansion of Green's  Function}
 This section of the appendix is devoted to proof of the expansion of the Green's matrix function.

 Let  $G(x,x')$  be a $n\times n$ matrix valued function which satisfies  following equation
 \begin{equation}\label{5}
 (-\frac{d^2}{d x^2}+E +V(x))G(x,x')=\delta(x-x')
 \end{equation}

 $V(x)=\theta (-x)  V + \theta (x) V_p$ , $V$ and $V_p$ are two real
 positive definite $n\times n$ matrices, and  generally noncommuting. The domain of the solution is $\mathbb{R}$.

 To construct $G$ we  shall use Weiner-Hopf method to turn the problem into inhomogeneous Hilbert problem and show that it can be obtained order by order around the $V(x)=0$ .

 We first reformulate (\ref{5}) and convert it into an integral equation.

To that end we introduce the free Green's function $G_0$ satisfying

 $$
 (-\frac{d^2}{d x^2}+E )G_0(x,x') = \delta(x-x')
 $$
 from which it follows that

 $$
 G_0(x,y) = \frac{1}{2\sqrt{E}}e^{-\sqrt{E}|x-y|}
 $$

 Using $G_0$ it is easy to show that $G(x,y)$ satisfies the following integral equation

 $$
 G(x,y)= G_0+\int_{-\infty}^{\infty} G_0(x,z)V(z) G(z,y) dz
 $$

 By iteration we obtain   the following expansion for $G$

 $$
 G(x,y)= \sum_{n=0}^{\infty} G_n(x,y)
 $$

 with $G_n(x,y)$ given by

 $$
 G_n(x,y)=(-1)^n \int dy_1  ..\int dy_n  G_0(x,y_1) G_0(y_1,y_2)...G_0(y_{n-1},y_n) G_0(y_n,y) V(y_1)...V(y_n)
 $$

 From which  the following recurrence relation follows

 \begin{equation}\label{4}
 G_n(x,y)=-\int_{-\infty}^{\infty} G_0(x,z)V(z)G_{n-1}(z,y)
 \end{equation}

 The ordering of all quantities appearing in the above equations and in  what follows is crucial,  as they are generally noncommuting.

 Let us construct the  following scaled matrix  function
 \begin{equation}\label{6}
 G_n^+(x,y)= \left\{
 \begin{array}{ll}
 0, & x >0; \\
 G_n(x,y), & x < 0.
 \end{array}
 \right.
 \end{equation}

 \begin{equation}\label{61}
 G_n^-(x,y)= \left\{
 \begin{array}{ll}
 -G_n(x,y), & x >0; \\
 0, & x < 0.
 \end{array}
 \right.
 \end{equation}

 Using (\ref{6}) and (\ref{61}) we can write (\ref{4}) as

 \begin{equation}\label{7}
 G_n^+(x,y)-G_n^-(x,y) =-V\int_{-\infty}^0dz G_0(x,z)G_{n-1}^+(z,y) +V_p\int_{-\infty}^0 dz G_0(x,z)G_{n-1}^-(z,y)
 \end{equation}

 By taking the two sided Laplace Transform of both sides of the above equation and using the fact that $G_0(x,y)$ depends only on the difference between $x$ and $y$,    the convolution theorem leads to the following recurrence relation  between a sequence of inhomogeneous Hilbert problems,

 \begin{equation}\label{8}
 F_n^+(s,y)-F_n^-(s,y) =-V F_0(s)F_{n-1}^+(s,y) +V_T F_0(s)F_{n-1}^-(s,y)= D_{n}(s,y)
 \end{equation}

 where $ F_n^{\pm}(s,y)$ and $F_0(s)$ are defined by

 $$
 F_n^{\pm}(s)= \int_{-\infty}^{\infty} e^{-sx}  G_n^{\pm}(x,y) dx
 $$

 Here we note that in view of the definition of $  G_n^{\pm}(x,y)$  we can infer that $ F_n^-(s)$ is a regular function ( analytic) for $Re (s) >0$ and $ F_n^{+}(s)$ regular for $Re(s) <0$.

 For $F_0(s)$ we have
 $$
 F_0(s)=\frac{1}{2}\int_{-\infty}^{\infty} e^{-sx-\sqrt{E}|x|} dx= \frac{1}{E-s^2}\mathbb{I}~~~~~~~~~~\text{regular inside the strip} ~~|Re(s)|<\sqrt{E}
 $$

 Now, equation (\ref{8}) is a discontinuity equation , the branch cut being the imaginary axis $Re(s)=0$,we can use standard theorems known for this kind of inhomogeneous Hilbert problem, to  solve these sequence  order by order to obtain $F_n(s)$,   from which one is computed $G_n(x,y)$ by inverting the Laplace transform.

 We shall work out explicitly the region $y<0,x<0$ up to the second order. The other regions are either similar or can be deduced simply from this region. The second order turned out to be enough for to guess the general solution.

 \begin{itemize}
 	\item{\bf{First Order}}
 	
 	$$
 	F_1^+(s,y)-F_1^-(s,y) =-V F_0(s)F_{0}^+(s,y) +V_p F_0(s)F_{0}^-(s,y)
 	$$
 	We have first to compute $F_{0}^{\pm}(s,y)$ .
 	
 	$$
 	F_{0}^-(s,y)= -\frac{1}{2m} \int_0^\infty e^{-s x} e^{-\sqrt{E}|x-y|} = -\frac{e^{\sqrt{E}y}}{2m(s+\sqrt{E})},~~~~~ Re(s) >0,  ~y,x<0
 	$$

 	For $ F_0^+(s,y) $
\begin{eqnarray}
G_{0}^+(s,y)&=& \frac{1}{2\sqrt{E}} \int_0^\infty e^{-s x} e^{-\sqrt{E}|x-y|}\nonumber \\
&=& -\frac{1}{2m}\big( \frac{e^{\sqrt{E}y}}{s+\sqrt{E}} +\frac{e^{-sy}}{s-\sqrt{E}} -\frac{e^{-sy}}{s-\sqrt{E}} \big) ~~,~~Re(s) <0 , y <0, x>0\nonumber
\end{eqnarray}

 	it is then straightforward to show that
 	
 	$$
 	D_1(x) =\delta V F_0F_0^- - V F_0^2 e^{-sy},~~ \delta V= V_p-V
 	$$
 	
 	from which we can compute $F_1^{\pm}(s,y)$
 	
 	For $F_1^+$
 	\be\label{disc1}
 	F_1^+(s,y) = \frac{1}{2\pi i}\int_{-i\infty}^{i \infty} \frac{D_1(z)}{z-s} dz = c_1 +c_2
 	\ee

 	$$
 	c_1= \frac{1}{2\pi i} \delta V \int_{-i\infty}^{i \infty} \frac{F_0 F_0^-}{z-s} dz
 	$$
 	
 	$$
 	c_2= -\frac{1}{2\pi i} V \int_{-i\infty}^{i \infty} \frac{F_0^2 e^{-zy}}{z-s} dz
 	$$
 	
 	We note that in general one has to add an arbitrary polynomial of $s$ ( matrix elements) to the above solution for $F_1^+$, however they are excluded by the requirement that $\lim_{s\rightarrow -\infty} F_1^+ = 0$.

 	$c_1$ and $c_2$ can be easily computed using contour integration.
 	
 	$$
 	c_1= \frac{\delta V}{(2\sqrt{E})^3} \frac{e^{\sqrt{E}y}}{s-\sqrt{E}}
 	$$

 	$$
 	c_2= - V \frac{e^{-sy}}{(s^2-E)^2} - V \frac{e^{\sqrt{E}y}}{4E} \bigg[ \frac{y}{\sqrt{E}+s}-\frac{1}{\sqrt{E}(\sqrt{E}+s)}-\frac{1}{(\sqrt{E}+s)^2}\bigg]
 	$$

 	therefore

 	$$
 	F_1^+= \frac{\delta V}{(2\sqrt{E})^3} \frac{e^{\sqrt{E}y}}{s-\sqrt{E}} - V \frac{e^{-sy}}{(s^2-E)^2} -V\frac{e^{\sqrt{E}y}}{4E}\bigg[ \frac{y}{\sqrt{E}+s}-\frac{1}{\sqrt{E}(\sqrt{E}+s)}-
 	\frac{1}{(\sqrt{E}+s)^2}\bigg]
 	$$
 	
 	$$
 	Re(s)<0
 	$$
 	For $F_1^-$ we can  use the same method or use directly the discontinuity equation to obtain for $F_1^+$
 	
 	$$
 	F_1^-= \frac{\delta V}{(2\sqrt{E})^3} \frac{e^{\sqrt{E}y}}{s-\sqrt{E}}\big[1-\frac{4E}{(s+\sqrt{E})^2}\big] - V \frac{e^{\sqrt{E}y}}{4E} \bigg[ \frac{y}{\sqrt{E}+s}-\frac{1}{\sqrt{E}(\sqrt{E}+s)}-\frac{1}{(\sqrt{E}+s)^2}\bigg]
 	$$
 	
 	$$
 Re(s)>0
 	$$
 	
 	Now,   we can compute $G_1(x,y)$ for $x, y<0$   using the inverse Laplace transform
 	
 	$$
 	G_1(x,y) =\frac{1}{2\pi i} \int_{-i\infty}^{i \infty} e^{sx} G_1^+(s,y) ds
 	$$
 	
 	and obtain
 	$$
 	G_1(x,y) = -\frac{\delta V}{(2\sqrt{E})^3} e^{\sqrt{E}(x+y)} -\frac{V}{2\sqrt{E}} G_0(x,y)\big ( |x-y|+\frac{1}{\sqrt{E}} \big )
 	$$

 	\item{\bf{Second Order}}
 	
 	$$
 	D_2=F_2^+-F_2^-=-VF_0F_1^++V_pF_0 F_1^-
 	$$
 	Writing  $F_1^+$ in terms of $F_1^-$ we get
 	
 	$$
 	D_2= \d V F_0 F_1^- +V^2 F_0^2 F_0^+ -VV_p F_0^2 F_0^-
 	$$
 	
 	Then
 	$$
 	F_2^+ (s,y)=\frac{1}{2\pi i}\int_{-i\infty}^{i\infty}  \bigg[\underbrace{\d V  G_0 G_1^-}_{1} +\underbrace{ V^2 G_0^2 G_0^+}_{2} - \underbrace{VV_p  G_0^2 G_0^-}_{3} \bigg] dz
 	$$
 	
 	The first term $(1)$ has two contributions
 	
 	\begin{eqnarray}
 	(1-1)&=&\frac{\d V^2}{(2\sqrt{E})^3} e^{\sqrt{E}y} \frac{1}{2\pi i }\int \frac{dz}{(E-z^2)(z-\sqrt{E})(z-s)} \big[ 1-(\frac{2\sqrt{E}}{\sqrt{E}+z})^2\big]   \\ \nonumber
 	&=& -\frac{1}{16\sqrt{E}}(\frac{\d V}{E})^2 \frac{e^{\sqrt{E}y}}{s-\sqrt{E}}
 	\end{eqnarray}

 	and
 	\begin{eqnarray}
 	(1-2)&=&\frac{\d V V}{4E} e^{\sqrt{E}y}\frac{1}{2\pi i } \int_{-i\infty}^{i\infty} \frac{dz}{(z^2-E)(z-s)} \big[ \frac{y}{\sqrt{E}+z}-\frac{1}{\sqrt{E}(\sqrt{E}+z)}\\ \nonumber&-& \frac{1}{(\sqrt{E}+z)^2}\big] dz
 	= \frac{1}{16}\frac{\d V V}{E^2}\frac{e^{\sqrt{E}y}}{s-\sqrt{E}} \big [
 	y -\frac{3}{2\sqrt{E}} \big ]
 	\end{eqnarray}
 	
 	To compute $(2)$ and $(3)$ we sum them and use the relation between $ F_0^+ $ and  $F_0^- $ to get
 	
 	$$
 	(2)+(3)= \underbrace{-V\d V\frac{1}{2\pi i } \int_{-i\infty}^{i\infty} G_0^2 G_1^-dz}_{A} +\underbrace{ V^2\frac{1}{2\pi i } \int_{-i\infty}^{i\infty} G_0^3 e^{-sy} dz}_{B}
 	$$
 	
 	$$
 	A=  \frac{1}{16}\frac{ V\d V }{E^2}\frac{e^{\sqrt{E}y}}{s-\sqrt{E}} \big [
 	\frac{1}{s-\sqrt{E}} -\frac{3}{2\sqrt{E}} \big ]
 	$$
 	
 	$$
 	B=-V^2\bigg [ \frac{e^{-sy}}{(s^2-E)^3} +\frac{e^{\sqrt{E}y}}{2 (2\sqrt{E})^3(\sqrt{E}+s)} \big( [y-\frac{3}{2\sqrt{E}}
 	-\frac{1}{\sqrt{E}+s}]^2 +\frac{3}{4E}+ \frac{1}{(s+\sqrt{E})^2}\big )\bigg ]
 	$$
 	
 	\begin{eqnarray}
 	G_2^+(s,y) &=&-\frac{1}{16\sqrt{E}}(\frac{\d V}{E})^2 \frac{e^{\sqrt{E}y}}{s-\sqrt{E}} -V^2 \frac{e^{-sy}}{(s^2-E)^3} \\ \nonumber
 	&+& \frac{1}{16}\frac{ V\d V }{E^2}\frac{e^{\sqrt{E}y}}{s-\sqrt{E}} \big [
 	\frac{1}{s-\sqrt{E}} -\frac{3}{2\sqrt{E}} \big ]+\frac{1}{16}\frac{\d V V}{E^2}\frac{e^{\sqrt{E}y}}{s-\sqrt{E}} \big [
 	y -\frac{3}{2\sqrt{E}} \big ] \\ \nonumber
 	&-& V^2 \frac{e^{\sqrt{E}y}}{2 (2\sqrt{E})^3(\sqrt{E}+s)} \big( [y-\frac{3}{2\sqrt{E}}
 	-\frac{1}{\sqrt{E}+s}]^2 +\frac{3}{(2\sqrt{E})^2}+ \frac{1}{(s+\sqrt{E})^2}\big )
 	\end{eqnarray}

 	$G_1(x,y)$ can then be computed easily by complex contour integration, with $x,y< 0$
 	\begin{eqnarray}
 	G_2(x,y) &=& \frac{1}{2\pi i} \int_{-i \infty }^{i\infty} e^{s x} F_1^+(s,x)  ds=  \frac{e^{-\sqrt{E}|x-y|}}{2\sqrt{E}} \bigg[  \frac{1}{16}(\frac{{V}}{E})^2\big[ (\sqrt{E}|x-y|+3/2)^2+3/4\big] \bigg]\nonumber \\
 	&+&	\frac{e^{\sqrt{E}(x+y)}}{2\sqrt{E}}\bigg[ \frac{1}{8}(\frac{\delta {V}}{E})^2-\frac{{V}\delta {V}}{16E^2} (\sqrt{E}x-3/2) -\frac{\delta {V}{ V}}{16E^2} (\sqrt{E}y-3/2)\bigg]
 	\end{eqnarray}

 	Collecting $G_1$ and $G_2$ it is trivial to recover the expansion of $G$ given by equation (\ref{GFapp}).
 \end{itemize}

 	\subsection{Continuity Conditions}
 	
 	The continuity of the eigenfunctions and their first derivatives at the cut $\t =0$ lead to the  following continuity  conditions for the Green's matrix function and its first derivatives.

 \begin{eqnarray}\label{continuity}
 \lim_{\t'\rightarrow 0} G_{-+}(\t,\t')&=& \lim_{\t'\rightarrow 0} G_{--}(\t,\t')\nonumber\\
 \lim_{\t'\rightarrow 0} G_{+-}(\t,\t')&=& \lim_{\t'\rightarrow 0} G_{++}(\t,\t') \nonumber\\
 \lim_{\t'\rightarrow 0}\partial_{\t}\ G_{-+}(\t,\t')&=& \lim_{\t'\rightarrow 0} \partial_{\t}\ G_{--}(\t,\t')\nonumber\\
 \lim_{\t\rightarrow 0}\partial_{\t}\ G_{-+}(\t,\t') &=& \lim_{\t\rightarrow 0} \partial_{\t}\ G_{++}(\t,\t') \nonumber\\
 \lim_{\t'\rightarrow 0}\partial_{\t}\ G_{+-}(\t,\t') &=& \lim_{\t'\rightarrow 0} \partial_{\t}\ G_{++}(\t,\t') \nonumber\\
 \lim_{\t\rightarrow 0}\partial_{\t}\ G_{+-}(\t,\t') &=& \lim_{\t\rightarrow 0} \partial_{\t}\ G_{--}(\t,\t')\nonumber \\
 \end{eqnarray}

 We have just listed half of the set of the resulting continuity equations, namely with respect to one argument( the first or the second), because the others follow from the hermiticity condition which we shall prove lastly.

 $ G_{\mp \mp}$ and $ {G}_{\pm\mp}$ are given by

 \be\label{GFSa}
 {G}_{\mp\mp}(\t,\t') = \frac{1}{2{W}_\mp} e^{-{W}_\mp|\t-\t'|}
 +\frac{1}{2{W}_\mp}  e^{\pm{W}_\mp\t}
 ({W}_\mp-{W}_\pm)({W}_-+{W}_+)^{-1}   e^{\pm{W}_\mp \t'}
 \ee

 and

 \be\label{GFSa1}
 {G}_{\pm\mp}(\t,\t')=  e^{\mp{W}_\pm\t}({{W}_-+{W}_+ })^{-1}
 e^{\pm{W}_\mp \t'}
 \ee

 $$
 \frac{1}{W_{\mp}} =W_{\mp}^{-1}
 $$

 We start by the first condition of equation (\ref{continuity}).
 $$
 \lim_{\t'\rightarrow 0} G_{-+}(\t,\t')= \lim_{\t'\rightarrow 0} G_{--}(\t,\t')
 $$
 \begin{eqnarray}
 \lim_{\t'\rightarrow 0} G_{-+}(\t,\t')&=&e^{W_-\t}({W_++W_-})^{-1} \nonumber\\
 \lim_{\t'\rightarrow 0} G_{--}(\t,\t') &=&\frac{1}{2W_-}.e^{W_-\t}+\frac{1}{2W_-}.e^{W_-\t}(W_--W_+)(W_++W_+)^{-1}\nonumber\\
 &=& e^{W_-\t}.\frac{1}{2W_-}.[\ 1+(W_--W_+)(W_++W_-)^{-1}\ ]\nonumber\\
 &=& e^{W_-\t}.\frac{1}{2W_-}.[\ (W_++W_-)+(W_--W_+)\ ](W_++W_-)^{-1}\nonumber\\
 &=&e^{W_-\t}(W_++W_-)^{-1}\nonumber
 \end{eqnarray}

 By exactly similar matrix manipulation we show.
 \begin{eqnarray}
 \lim_{\t'\rightarrow 0} G_{+-}(\t,\t')&=& \lim_{\t'\rightarrow 0} G_{++}(\t,\t'))\nonumber
 \end{eqnarray}

 Consider now the continuity of the derivatives.

 First it is easy to show the following equalities.
 \begin{eqnarray}
 \partial_{\t}\ G_{++}(\t,\t')&=& -\frac{\mid \t-\t'\mid}{2(\t-\t')}e^{-W_+\mid \t-\t'\mid}-\frac{1}{2}.e^{-W_+\t}(W_+-W_-)(W_++W_-)^{-1}.e^{-W_+\t'} \nonumber\\
 \partial_{\t}\ G_{--}(\t,\t')&=&-\frac{\mid \t-\t'\mid}{2(\t-\t')}e^{-W_-\mid \t-\t'\mid}+\frac{1}{2}.e^{W_-\t}(W_+-W_-)(W_++W_-)^{-1}.e^{W_-\t'} \nonumber\\
 \partial_{\t}\ G_{+-}(\t,\t')&=&-W_+.e^{-W_+\t}(W_++W_-)^{-1}.e^{W_-\t'} \nonumber\\
 \partial_{\t}\ G_{-+}(\t,\t')&=& W_-.e^{w_-\t}(W_++W_-)^{-1}.e^{-W_+\t'} \nonumber
 \end{eqnarray}

 We shall work out one conditions explicitly, the others are very similar as can easily be noticed.

 We take
 \begin{eqnarray}
 \lim_{\t'\rightarrow 0}\partial_{\t}\ G_{-+}(\t,\t')&=& \lim_{\t'\rightarrow 0} \partial_{\t} G_{--}(\t,\t')\nonumber\\
 \lim_{\t'\rightarrow 0}\partial_{\t}\ G_{-+}(\t,\t')&=& W_-.e^{W_-\t}(W_++W_-)^{-1} \nonumber\\	
 \lim_{\t'\rightarrow 0}\partial_{\t} G_{--}(\t,\t') &=&\frac{1}{2}e^{W_-\t}+\frac{1}{2}.e^{W_-\t}(W_--W_+)(W_++W_-)^{-1} \nonumber \\
 &=& \frac{1}{2} e^{W_-\t}[(1+(W_--W_+)(W_++W_-)^{-1}] \nonumber\\
 &=& W_-.e^{W_-\t}(W_++W_-)^{-1}\nonumber
 \end{eqnarray}

 Let us finally check the hermiticity constraint.

 The Green's matrix function solution must by construction satisfy the following  identity
 $$	
 G(\t,\t')^{\dagger} = G(\t',\t)^{\dagger}
 $$
 this leads to the following conditions
 \be\label{const1}
 G_{-+}(\t,\t')^{\dagger} = G_{+-}(\t,\t')
 \ee

 \be\label{const2}
 G_{\pm \pm}(\t,\t')^{\dagger} = G_{\pm \pm}(\t,\t')
 \ee	

 The first constraint, equation (\ref{const1}) is easily seen to be satisfied. However the two constraints given by  equation (\ref{const2}), are not obvious, therefore we shall work out one of them explicitly.

 $G_{++}(\t,\t')^{\dagger} = G_{++}(\t',\t)$.

 \begin{eqnarray}		
 G_{++}(\t,\t')^{\dagger}=  \frac{1}{2{W}_+} e^{-{W}_+|\t-\t'|}
 +\frac{1}{2}  e^{-{W}_+\t'}
 ({W}_-+{W}_+)^{-1} ({W}_--{W}_+)  {W}_+^{-1}e^{-{W}_+ \t}
 \end{eqnarray}

 Therefore we need to show that
 $$
 ({W}_-+{W}_+)^{-1} ({W}_--{W}_+)  {W}_+^{-1}={W}_+^{-1} ({W}_--{W}_+) ({W}_-+{W}_+)^{-1}
 $$
 \begin{eqnarray}
 ({W}_-+{W}_+)^{-1} ({W}_--{W}_+)  {W}_+^{-1}&=& 	({W}_-+{W}_+)^{-1}{W}_-{W}_+^{-1}-({W}_-+{W}_+)^{-1} \nonumber \\
 &=&W_+^{-1} W_{+}({W}_-+{W}_+)^{-1}{W}_-{W}_+^{-1}-({W}_-+{W}_+)^{-1}\nonumber\\
 &=& W_+^{-1} W_{-}({W}_-+{W}_+)^{-1}-({W}_-+{W}_+)^{-1}\nonumber \\
 &=&  W_+^{-1} W_{-}({W}_-+{W}_+)^{-1}- W_+^{-1}  W_+({W}_-+{W}_+)^{-1} \nonumber \\
 &=& {W}_+^{-1} ({W}_--{W}_+) ({W}_-+{W}_+)^{-1}\nonumber 	
 \end{eqnarray}

 where we have used  the famous matrix identity
  $$
 W_+ (W_++W_-)^{-1}.W_-= 	W_- (W_++W_-)^{-1} W_+
 $$
 	
 	to go from the second to the third line.
 	\section{Appendix II}

 		The aim is to compute the following limit
 	
 	\begin{equation}\label{traceanzats}
 	\lim_{N\rightarrow \infty} \frac{\mathrm{Tr}M_N(\alpha)}{N}= \lim_{N\rightarrow \infty}\frac{1}{N^2}\sum_{k,p=0}^{N-1}\sum_{l=-(N-1)}^{N-1} (1-\frac{|l|}{N}) \alpha_{kp} e^{i2\pi(p-k+1/2)l/N}
 	\end{equation}
 	
 	First it is easy to show that for arbitrary $N$  and $\alpha_{kp}$ we have
 	
 	\begin{eqnarray}
 	\sum_{l=1}^{N-1}\alpha_{kp}e^{2\pi i l(p-k+1/2)/N}+\sum_{l=1}^{N-1}\alpha_{kp}e^{2\pi i (-l).(p-k+1/2)/N}=0
 	\end{eqnarray}
 	
 	Therefore we are left with three contributions.
 	$$
 	I_0= \sum_{k,p=0}^{N-1}\alpha_{kp}
 	$$
 	
 	$$
 	I_-=(I_+)^* = \sum_{k,p=0,l=1}^{N-1}\ \frac{l}{N^3}.\alpha_{kp}e^{\frac{2\pi i l}{N}(p-k+1/2)}
 	$$
 	
 	It will be shown that $I_-$ is real in the limit $N\rightarrow \infty$ and   therefore  $I_-$ and $I_+$ give equal contributions.
 	
 	We start by evaluating $I_0$.

 	\begin{eqnarray}
 	I_= \lim_{N\rightarrow \infty}	\frac{1}{N^2}\sum_{k,p=0}^{N-1}\alpha_{kp}&=&\frac{1}{2\pi i.N}\sum_{k=0}^{N-1}\oint_{|z|=1} \frac{dz}{z}\frac{1}{\alpha_+\sqrt{c_k}+\alpha_-\sqrt{R_z}}
 	\end{eqnarray}

 	Using complex contour integration, it is easy to convert the  above contour integral into the following integral

 	\begin{equation}
 	I_0= \frac{1}{\pi N}\sum_{k=0}^{N-1}\int_0^{r_-}\frac{dx}{x}\frac{\alpha_-\sqrt{R_x}}{\alpha_+^2 c_k+\alpha_-^2 R_x}
 	\end{equation}

 	where $R_x= (r_+-x)(r_--x)/2x$ .  $ r_+,r_-$ are in addition to $z=0$ the branch points of the function $\sqrt{R_z}= \sqrt{\frac{2az-z^2-1}{2z}}$.
 	$$
 	r_- =a-\sqrt{a^2-1},~~~r_+=1/r_-
 	$$
 	
 	then we have

 	$$
 	I_0	=\frac{\alpha_-}{\pi(2\pi i)}\int_0^{r_-} dx\frac{\sqrt{R_x}}{x}\oint_{|z|=1} \frac{dz}{z}\frac{1}{\alpha_+^2.R_z+\alpha_-^2R_x}=\frac{\alpha_-}{\pi }\int_0^{r_-} \frac{\sqrt{R_x}}{\alpha_+^2x\sqrt{e^2-1}}dx
 	$$
 	
 	where $e=a+\frac{\alpha_-^2}{\alpha_+^2}	 R_x$.

 	Consider now $I_-$.

 	\begin{eqnarray}\label{Iminus}
 	I_-=\frac{1}{2\pi i.N^2}\sum_{k=0,l=1}^{N-1}\ l.e^{\frac{2\pi i l}{N}(-k+1/2)}\oint_{|z|=1} dz \left[\frac{z^{l-1}}{\alpha_+\sqrt{c_k}+\alpha_-\sqrt{R_z}}+\frac{z^{l-1-N}}{\alpha_+\sqrt{c_k}+\alpha_-\sqrt{R_z}}\right]
 	\end{eqnarray}
 	As can be noticed in converting the sum of $p$ into an integral we have added another term proportional to $z^{l-1-N}$  which comes from the reminder when using Euler-Maclaurin, this term can not be neglected within this sum and actually it is this term which survives at the end.
 	
 	Now, using again complex contour integration it is easy to show that

 	\begin{eqnarray}
 	\sum_{l=1}^{N-1}e^{\frac{2\pi i l}{N}(-k+1/2)}\frac{l}{N}\frac{z^{l-1}}{\alpha_+\sqrt{c_k}+\alpha_-\sqrt{R_z}}=\sum_{l=1}^{N-1}e^{\frac{2\pi i l}{N}(-k+1/2)}\frac{l}{N}\int_0^{r_-}dx.x^{l-1}\frac{2i\alpha_-\sqrt{R_x}}{\alpha_+^2c_k+\alpha_-^2 R_x}\nonumber
 	\end{eqnarray}
 	
 	which is easily seen to vanish in the limit $N\rightarrow \infty$, in view of the fact that $ x<r_-<1$. Therefore we are left only with the second term of
 	equation (\ref{Iminus}). This term after summing over $l$ gives

 	\begin{eqnarray}
 	I_-&=&\frac{\alpha_-}{\pi N}\int_0^{r_-}dx\ \sqrt{R_x}\sum_{k=0}^{N-1}\left(\frac{1}{x-e^{\frac{2\pi i}{N}k}}.\frac{1}{\alpha_+^2.c_k+\alpha_-^2.R_x}\right)^*\nonumber\\
 	&=& \frac{\alpha_-}{\pi}\int_0^{r_-}dx\ \sqrt{R_x}\left(\frac{1}{2\pi i} \oint_{|z|=1} \frac{dz}{z} .\frac{1}{(x-z).(\alpha_+^2.R_z+\alpha_-^2.R_x)} \right)^*\nonumber\\
 	&=& \frac{\alpha_-}{\pi}\int_0^{r_-}dx\ \sqrt{R_x}\left( -\frac{1}{x(\alpha_-^2-\alpha_+^2)R_x}+\frac{1}{\alpha_+^2.(x-u)\sqrt{e^2-1}}\right)= I_+
 	\end{eqnarray}
 	
 	Where we have neglected all terms which vanish in the limit $N\rightarrow \infty$.
 	
 	Where $u=e-\sqrt{{e}^2-1}$ is a pole beside $x$ lying inside the unit circle.

 	Putting everything together we find
 	
 	\begin{equation}
 	\lim_{N\rightarrow \infty} \frac{\mathrm{Tr}M_N(\alpha)}{N}=\frac{\alpha_-}{\pi(\alpha_-^2-\alpha_+^2)}\int_{0}^{r_-}\big[ \frac{\frac{dR_x}{dx}}{\sqrt{e^2-1}\sqrt{R_x} } +\frac{1}{x \sqrt{R_x}}\big]dx
 	\end{equation}

 	It is now a matter of trvial exercise to show using Integral Tables that
 	\begin{equation}
 	\lim_{N\rightarrow \infty} \frac{\mathrm{Tr}M_N(\alpha)}{N}=\frac{1}{\pi}\frac{1}{\sqrt{a+1}}.K\left(\sqrt{\frac{2}{a+1}}\right)
 	\end{equation}
 	
 	We note that in all the above calculation we assumed $\alpha_+\neq \alpha_-\neq 1$. This particular case has to be worked out separably although the final result does not depend on the particular values of $\alpha_+$ and $ \alpha_-$, as long as they sum up to $2$.
 	
 	\section{Appendix III}
 	
 	This appendix is to dedicated  to the proof of  equation (\ref{corr1}) of section.

 	The main task is to compute the diagonal elements of the matrix $\mathbb{G}$, in order to
 	compute the first correction expression :
 	\begin{eqnarray}
 	\int_{-\infty}^{\infty} [\sum_{i=1}^{2n}(\mathbb{G})_{ii}^2(\tau,\tau)-n\sum_{i=1}^{2}(G_{ii})^2(\tau,\tau) ] d\tau
 	\end{eqnarray}
 	
 	Let us note that all quantities here are evaluated at $E=0$.
 	
 	For ($\tau<0$) the Green's matrix function is given by :
 	\be
 	\mathbb{G}(\tau,\tau)_{--}=\frac{1}{2\mathbb{W}}+\frac{1}{2\mathbb{W}}e^{\mathbb{W}\tau}(\mathbb{W}-\mathbb{W}_p)(\mathbb{W}+\mathbb{W}_p)^{-1}e^{\mathbb{W}\tau}
 	\ee
 	
 	It turns out that it is technically more convenient to write $\mathbb{G}_{--}(\tau,\tau)$ as
 	
 	\be\label{Grexp}
 	\mathbb{G}_{--}(\tau,\tau)=\frac{1}{2\mathbb{W}}-\frac{e^{2\mathbb{W}\tau}}{2\mathbb{W}}+e^{\mathbb{W}\tau}({\mathbb{W}+\mathbb{W}_p})^{-1}e^{\mathbb{W}\tau}
 	\ee
 	First it is easy to show that
 	
 	\begin{eqnarray}
 	\mathbb{W}_{ii}^{-1}&=&\frac{1}{2\sqrt{b}}(\lambda^{-1}_-+\lambda^{-1}_+)
 	\end{eqnarray}
 	
 	With $ \lambda_{\pm}=\sqrt{a\pm1}$
 	
 	and similarly for
 	
 	\be
 	(\frac{e^{2\mathbb{W}.\tau}}{2\mathbb{W}})_{ii}=
 	\frac{1}{4\sqrt{b}}(\frac{e^{2\sqrt{b}.\lambda_+\tau}}{\lambda_+}+\frac{e^{2\sqrt{b}.\lambda_-\tau}}{\lambda_-})
 	\ee
 	
 	The less straightforward piece is the last term  of eqt(\ref{Grexp}).  To compute its diagonal elements we first introduce the following orthogonal $2\times 2$ matrix and its $n$-fold version.
 	
 	$$
 	T= \begin{bmatrix}
 	\frac{1}{\sqrt{2}} & \frac{1}{\sqrt{2}}  \\
 	\frac{1}{\sqrt{2}} & - \frac{1}{\sqrt{2}} \\
 	\end{bmatrix}, ~~ \mathbb{T}= I_n \otimes T
 	$$
 	
 	$$
 	W=T.W_d.T,~~~~ \mathbb{ W}=\mathbb{T} (  I_n  \otimes W_d) \mathbb{T}
 	$$
 	
 	$$
 	W_d=\sqrt{b}
 	\begin{bmatrix}
 	\lambda_+ & 0  \\
 	0 & \lambda_- \\
 	\end{bmatrix}
 	$$
 	
 	Now, it is easy to observe that $\mathbb{T}.(\mathbb{W}+\mathbb{W}_p).\mathbb{T}$ is a  block circulant matrix, we call it $\mathbb{C}$, with block elements given by $(C_0,C_1,0,...,0,C_{n-1})$.
 	
 	Where

 	\begin{eqnarray}
 	C_0=\sqrt{b}\begin{bmatrix}
 	2\rho+\sigma & 0  \\
 	0 & 2\rho-\sigma \\
 	\end{bmatrix},\ \ \ C_1=\frac{\sqrt{b}\sigma}{2}\begin{bmatrix}
 	1 & -1  \\
 	1 & -1 \\
 	\end{bmatrix},\ \ \ C_{n-1}=\frac{\sqrt{b}\sigma}{2}\begin{bmatrix}
 	1 & 1  \\
 	-1 & -1 \\
 	\end{bmatrix}
 	\end{eqnarray}
 	
 	$$
 	\rho = \frac{\l_++\l_-}{2}, ~~~~~~ \sigma= \frac{\l_+-\l_-}{2}
 	$$
 	The block eigenvalues of the matrix $\mathbb{C}$ are :
 	\begin{eqnarray}
 	B_j=\sqrt{b}\begin{bmatrix}
 	2\rho+\sigma(1+\cos(2\pi j/n)\ ) & i\sigma\sin(2\pi j/n)  \\
 	-i\sigma\sin(2\pi j/n) & 2\rho-\sigma(1+\cos(2\pi j/n)) \\
 	\end{bmatrix}\ \ \ \ \ j=0..n-1
 	\end{eqnarray}
 	The inverses of \{$B_j$\} are needed for later use and given by :
 	\begin{eqnarray}
 	B^{-1}_j=\frac{b^{-1/2}}{(2\rho)^2-2\sigma^2(1+cos(2\pi j/n)\ )}\begin{bmatrix}
 	2\rho-\sigma(1+\cos(2\pi j/n)\ ) & -i\sigma\sin(2\pi j/n)  \\
 	i\rho\sin(2\pi j/n) & 2\rho+\sigma(1+\cos(2\pi j/n)) \\
 	\end{bmatrix}\nonumber
 	\end{eqnarray}
 	
 	Now we are ready to compute $ (e^{\mathbb{W}\tau}\frac{1}{\mathbb{W}+\mathbb{W}_p}.e^{\mathbb{W}\tau})_{ii}$

 	Let $\mathbb{E}_i$ denote the standard basis, $(\mathbb{E}_i)_j =\delta_{ij}, ~~~i,j=0,1,\cdots 2n-1$.
 	
 	\begin{eqnarray}\label{lastterm}
 	(e^{\mathbb{W}\tau}\frac{1}{\mathbb{W}+\mathbb{W}_p}.e^{\mathbb{W}\tau})_{ii}&=& \mathbb{E}_i^T \mathbb{T} e^{W_d.\tau}.\mathbb{C}^{-1}.e^{W_d.\tau}\mathbb{T}\mathbb{E}_i \nonumber \\
 	&=& \mathbb{E}_i^T \mathbb{T} e^{W_d.\tau}.\mathbb{F}^{-1}.\mathbb{C}_d^{-1}.\mathbb{F}.e^{W_d.\tau}\mathbb{T}  \mathbb{E}_i \nonumber\\
 	&=& \mathbb{E}_i^T \mathbb{T}.\mathbb{F}^{-1} e^{W_d.\tau}.\mathbb{C}_d^{-1}.e^{W_d.\tau}.\mathbb{F}\mathbb{T}  \mathbb{E}_i\nonumber
 	\end{eqnarray}
 	Where $\mathbb{F}$ is the discrete -block- Fourier transform.

 	It is convinient to write $ \mathbb{E}_i$ as
 	
 	$$
 	\mathbb{E}_l^m= u_m\otimes e_l ,~~~~~u_1= \left(
 	\begin{array}{c}
 	1\\
 	0
 	\end{array}
 	\right),~~~u_2= \left(
 	\begin{array}{c}
 	0\\
 	1
 	\end{array}
 	\right)
 	, ~~(e_l)_j=\delta_{lj},~~l,j=0,1\cdots n-1
 	$$
 	
 	now it is not difficult to show that
 	\begin{eqnarray}
 	\mathbb{F}\mathbb{T}\mathbb{E}_l^m=T.u_m\otimes\frac{1}{\sqrt{n}}
 	\left(
 	\begin{array}{c}
 	1\\
 	w^l\\
 	w^{2l}\\
 	.\\
 	.\\
 	.\\
 	.\\
 	w^{l(n-1)}
 	\end{array}
 	\right)
 	,~~~~  w= e^{2\pi i/n}
 	\end{eqnarray}
 	
 	Actually the result will be independent of $i$ or $l,m$.
 	
 	Let  $M_j= e^{W_d\tau}.B_j^{-1}.e^{W_d\tau}$, then we obtain from equation (\ref{lastterm})
 	\begin{eqnarray}
 	(e^{\mathbb{W}\tau}\frac{1}{\mathbb{W}+\mathbb{W}_p}.e^{\mathbb{W}\tau})_{ii}&=&\frac{1}{n}(T.u_m)^T.\sum_{j=0}^{n-1}M_j.T.u_m\nonumber\\
 	&=&\sum_{j=0}^{n-1}\frac{1}{2n}[ e^{2\sqrt{b}.\lambda_+\tau}(B^{-1}_j)_{11}+e^{2\sqrt{b}.\lambda_-\tau}(B^{-1}_j)_{22}+\nonumber\\
 	&-&(-1)^m.e^{\sqrt{b}.(\lambda_-+\lambda_+)\tau}((B^{-1}_j)_{12}+(B^{-1}_j)_{21})]\nonumber\\
 	&=&\frac{1}{2n}[e^{2\sqrt{b}.\lambda_+\tau}B_++e^{2\sqrt{b}.\lambda_-\tau}B_-]\nonumber
 	\end{eqnarray}
 	Where $B_+=\sum_{j=0}^{n-1}(B_j^{-1})_{11},\ B_-=\sum_{j=0}^{n-1}(B_j)^{-1}_{22}$. The terms $(B_j^{-1})_{21}+(B_j)^{-1}_{12}$ cancel out.
 	
 	Then it follows that $\mathbb{G}_{ii}$ for $(\tau<0)$ is given by
 	\begin{eqnarray}
 	(\mathbb{G}_{--}(\tau,\tau))_{ii}&=&
 	\frac{1}{4\sqrt{b}}[\ \lambda^{-1}_++\lambda^{-1}_-+\frac{e^{2\lambda_-\tau}}{\lambda_-}+\frac{e^{2\lambda_+\tau}}{\lambda_+}+\frac{2\sqrt{b}}{n}(e^{2\lambda_+\tau}B_++e^{2\lambda_-\tau}B_-)\ ]\nonumber
 	\end{eqnarray}
 	
 	For $(\tau>0)$ or $ \mathbb{G}_{++}(\tau,\tau))_{ii}$ one needs not to repeat the calculation, because  it is easy to observe using the action of the permutation matrix   that positive and negative times have equal contribution. Therefore we obtain

 	\begin{eqnarray}
 	\int_{-\infty}^{\infty}d\tau \sum_{i=1}^{2n}(\mathbb{G})_{ii}^2(\tau,\tau)-n.\sum_{i=1}^{2}(G_{ii})^2(\tau,\tau)=2\int_{-\infty}^{0}d\tau \sum_{i=1}^{2n}(\mathbb{G})_{ii}^2(\tau,\tau)-n.\sum_{i=1}^{2}(G_{ii})^2(\tau,\tau)\nonumber
 	\end{eqnarray}
 	Where $G=(2W)^{-1}$, which means that $G_{ii}^2=(\lambda_-^{-1}+\lambda_+^{-1})^2/(16b)\ $.
 	
 	For $ (\mathbb{G}_{ii})^2$ we have
 	\begin{eqnarray}
 	(\mathbb{G}_{ii})^2&=&\frac{1}{16b}\bigg[ \lambda_+^{-2}+\lambda_-^{-2}+2\lambda_+^{-1}.\lambda_-^{-1}+\frac{e^{4\sqrt{b}\lambda_+\tau}}{\lambda_+^2}+\frac{e^{4\sqrt{b}\lambda_-\tau}}{\lambda_-^2}+\frac{2e^{2\sqrt{b}(\lambda_-+\lambda_+)\tau}}{\lambda_+.\lambda_-}+\nonumber\\
 	&-& 2(\lambda_+^{-1}+\lambda_-^{-1})(\frac{e^{2\sqrt{b}\lambda_+\tau}}{\lambda_+}+\frac{e^{2\sqrt{b}\lambda_-\tau}}{\lambda_-})+\frac{2b}{n^2}(\ e^{4\sqrt{b}\lambda_+\tau}B_+^2+\nonumber\\
 	&+& e^{4\sqrt{b}\lambda_-\tau}B_-^2+2e^{2\sqrt{b}(\lambda_-+\lambda_+)\tau}B_-.B_+)+\frac{2\sqrt{b}}{n}(\lambda_+^{-1}+\lambda_-^{-1}+\nonumber\\
 	&-&\lambda_-^{-1}e^{2\sqrt{b}\lambda_-\tau}-\lambda_+^{-1}e^{2\sqrt{b}\lambda_+\tau}).(\ e^{2\sqrt{b}\lambda_+\tau}B_++(e^{2\sqrt{b}\lambda_-\tau})B_-\ )\bigg ]\nonumber
 	\end{eqnarray}
 	
 	The integration and summation are now straightforward  :
 	\begin{eqnarray}
 	\int_{-\infty}^{\infty}d\tau \sum_{i=1}^{2n}(\mathbb{G})_{ii}^2(\tau,\tau -n\sum_{i=1}^{2}(G_{ii})^2(\tau,\tau)&=&\frac{1}{b\sqrt{b}}\bigg[n(-\frac{3}{16\lambda_+\lambda-}-\frac{3}{16\lambda_+^2}-\frac{1}{4\lambda_+^2\lambda_-}\nonumber\\-\frac{1}{4\lambda_+\lambda_-^2}
 	+\frac{1}{4\lambda_-\lambda_+(\lambda_-+\lambda_+)})&+& (\frac{1}{2\lambda_+\lambda_-}+\frac{1}{4\lambda_+^2}-\frac{1}{2\lambda_-(\lambda_++\lambda_-)})B_-\nonumber\\
 	+(\frac{1}{2\lambda_+\lambda_-}+\frac{1}{4\lambda_-^2}-\frac{1}{2\lambda_+(\lambda_++\lambda_-)})B_+ &+&\frac{1}{n}(\frac{1}{4\lambda_+}B_-^2+\frac{1}{4\lambda_-}B_+^2+\frac{1}{\lambda_-+\lambda_+}B_-B_+))\bigg]\nonumber
 	\end{eqnarray}

\bibliographystyle{JHEP}
\bibliography{document.bib}
\end{document}